\documentclass[12pt]{article}
\usepackage{epsfig}
\def\be{\begin{eqnarray}}
\def\ee{\end{eqnarray}}

\def\la{\langle}

\newcommand{\al}{\alpha}

\newcommand{\dl}{\delta}

\newcommand{\eq}{\begin{equation}}
\newcommand{\eqx}{\end{equation}}
\newcommand{\eqn}{\begin{eqnarray}}
\newcommand{\eqnx}{\end{eqnarray}}
\newcommand{\ben}{\begin{eqnaray}}
\newcommand{\een}{\end{eqnarray}}
\newcommand{\f}[2]{\frac{#1}{#2}}

\newcommand{\GG}{{\cal G}}

\newcommand{\MM}{{\cal M}}
\newcommand{\BB}{{\cal B}}
\newcommand{\ZZ}{{\cal Z}}

\newcommand{\HH}{{\cal H}}

\newcommand{\zb}{\bar{z}}

\newcommand{\arr}[4]{
\left(\begin{array}{cc}
#1&#2\\
#3&#4
\end{array}\right)
}

\newcommand{\cor}[1]{\left\langle{#1}\right\rangle}
\newcommand{\ket}[1]{\left|{#1}\right\rangle}

\newcommand{\tr}{\mbox{\rm tr}\,}

\newcommand{\lm}{\lambda}

\newcommand{\qqqq}{\quad\quad\quad\quad}

\begin{document}

%\begin{frontmatter}

\title{Infinite Products of Large Random Matrices and Matrix-valued
Diffusion}
%\author[Jagel,GSI]{Ewa Gudowska-Nowak},
%\author[Jagel]{Romuald A. Janik},
%\author[Jagel]{Jerzy Jurkiewicz} and
%\author[Jagel,GSI]{Maciej A. Nowak}

\author{Ewa Gudowska-Nowak$^{a,b}$\footnote{e-mail: {\tt gudowska@th.if.uj.edu.pl}},
Romuald A. Janik$^b$\footnote{e-mail: {\tt ufrjanik@if.uj.edu.pl}},\\  
Jerzy Jurkiewicz$^b$\footnote{e-mail: {\tt jurkiewicz@th.if.uj.edu.pl}} and Maciej
A. Nowak$^{a,b}\footnote{e-mail: {\tt nowak@th.if.uj.edu.pl}}$\\ 
$^a$ \small Gesellschaft f\"{u}r Schwerionenforschung,\\
\small Planckstrasse~1,\\
\small D-64291 Darmstadt,  Germany\\
$^b$ \small M. Smoluchowski Institute  of Physics,\\ 
\small Jagellonian University,\\ 
\small Reymonta 4,\\
\small PL 30-059 Krak\'{o}w, Poland}

%\address[Jagel]{M. Smoluchowski Institute  of Physics, 
%Jagellonian University, Reymonta 4, PL 30-059
%        Krak\'{o}w,Poland}
%\address[GSI]{Gesellschaft f\"{u}r Schwerionenforschung,
%D-64291 Darmstadt, Planckstrasse~1, Germany.}

\date{\today}

\maketitle

\begin{abstract}
We use  an extension of the diagrammatic rules in random matrix 
theory to evaluate spectral properties of finite and infinite products
of large complex matrices  and  large 
hermitian  matrices. The infinite product case allows us to define
a natural matrix-valued multiplicative diffusion process.
In both cases of hermitian and complex matrices, 
we observe the emergence  of a "topological phase transition", when a
hole develops in the eigenvalue spectrum, after some critical
diffusion time $\tau_{\rm crit}$ is reached.
In the case of a particular product of two hermitian ensembles, 
we observe also an unusual localization-delocalization phase transition
in the spectrum of the considered ensemble. We verify the analytical
formulas obtained in this work by numerical simulation. 
\\[4mm]
\end{abstract}

%\end{frontmatter}

%%%%%%%%%%%%%%%%%%%%%%%%

% \eject
% \newlength{\oldparskip}
% \setlength{\oldparskip}{\parskip}
% \addtolength{\parskip}{-4pt}
% \tableofcontents
% \vfil
% \eject
% \parskip \oldparskip

\pagebreak

\tableofcontents

\section{Introduction}

Applications of random matrix theory (RMT) in physics range
from interpretation of complex spectra of energy levels in atomic and
nuclear physics~\cite{GENERALHER}, studies of disordered 
systems~\cite{MONTAMBAUX,GUHR}, chaotic behavior~\cite{HAAKE} to quantum 
gravity~\cite{AMBJORN,AMBJORNETAL,ZINNJUSTIN}. 
Use of progressively developed methods of  RMT has also proved to be of
particular interest in  other sciences such as meteorology \cite{SAN},
image processing \cite{SET}, population ecology \cite{CASWELL} or
economy~\cite{MANTEGNA,BOUCHAUD}.

In this work, we will be concerned with infinite products of random 
matrices (PRM) sampled from general Gaussian ensembles.
Products of that type  appear in various branches of physics
and interdisciplinary applications \cite{VULPIANI} where system
dynamics is described in terms of 
evolution operators with random coefficients.  
Perhaps the best known examples are those related to thermal properties of
disordered magnetic systems, in particular, to the 
localization of electronic wave 
functions in random potentials \cite{BEENAKER,MONTAMBAUX}. 
Attention has  also been focused on applications of PRM
to the analysis of chaotic dynamical systems
\cite{VULPIANI,WANG,DIAKONOS,PETRI},  where stability of  
trajectories and their sensitive dependence on initial conditions are
measured in terms of characteristic Lyapunov exponents.
Moreover, there are  several results exploring the use of  PRM
formalism in applied physics and interdisciplinary  
research, ranging from the studies of  stability of large eco-
and social- systems \cite{CASWELL}, adaptive algorithms and
the analysis of system performance under the influence  
of external noises \cite{TSE} to image compression \cite{BARNSLEY}
and communication via antenna arrays \cite{TELETAR,MULLER}.
However, despite  the richness of the potential applications, the number of 
analytical results available in the PRM theory is still rather limited. 

In this paper we develop effective calculational techniques for
studying products of random matrices in the large $N$ limit, and use
these tools to derive the properties of the natural matrix valued
generalization of the geometric (multiplicative) diffusion type
process. The scalar versions of these processes, leading to log-normal
distributions are ubiquitous in various fields~\cite{ONEDIMREV} and we
expect the matrix valued generalizations to find numerous applications.

Let us briefly review the conventional (scalar) multiplicative
(geometric) random walk in one dimension in the presence of some
external drift force. The process belongs to the class of  (Markovian)
Ito diffusion processes, whose stochastic variable $s$ undergoes an evolution
\be
\frac{ds}{s} = \mu(s,t) dt + \sigma(s,t) dW_t 
\label{ito}
\ee
In the above  stochastic differential equation (SDE), part of the 
evolution is driven 
by a variable $dW_t$ that represents the Wiener process (integral of
the Gaussian white-noise), respecting
\be
\cor{dW_t}=0 \,\,\,\,\,\,\,\,\,\,\, \cor{dW_t^2}=dt
\ee
For simplicity we will limit ourselves to the constant drift $\mu$  and 
constant variance $\sigma$. 
Finite time evolution of the system  could be viewed as a string of increments 
\be
s(T)= \lim_{M \to \infty} \left [ (1+\mu \frac{T}{M}
+\sigma\sqrt{\frac{T}{M}}x_1 )(1+\mu \frac{T}{M} + 
\sigma\sqrt{\frac{T}{M}}x_2)\ldots\nonumber \right . \\
 \left . \ldots (1+\mu \frac{T}{M} +\sigma \sqrt{\frac{T}{M}}x_M)
\right ]s(0) %\nonumber \\
%=s(0)e^{\mu-\frac{\sigma^2}{2}t+\sigma W_t}
\label{onedimprod}
\ee
where $M$ stands for the number of steps in the discretization of the
time interval.
After averaging $s(T)$ over the independent identical distributions (iid)
of the Gaussian variables $x_i$ and taking the limit $M \to \infty$,
 we recover\footnote{The log-normal distribution $s(T)$ is easy to
infer looking at the  
form of the product (\ref{onedimprod}). Taking the logarithm, 
expanding and using the central limit theorem we immediately see, that 
the r.h.s. tends to the Gaussian distribution.}  
the well known solution for the probability density of $s(T)$:
\be
p(s,T|s_0,0)=\frac{1}{ s\sqrt{2\pi \sigma^2T}}\exp\left[-\frac{(\log(s/s_0)-\mu T+\frac{1}{2}\sigma^2
T)^2}{2\sigma^2 T}\right]
\label{lognormal}
\ee
with $s$ restricted to positive values. 

The aim of this paper is to study a similar construction in the space
of matrices taking the defining equation (\ref{onedimprod}) as a guideline
for the generalization.
We are therefore interested in properties of the
{\it matrix-valued evolution operator} defined as  
\be
Y(T)= \lim_{M \to \infty}  \left [ (1+\mu \frac{T}{M} +\sigma
\sqrt{\frac{T}{M}}X_1) 
(1+\mu \frac{T}{M} +\sigma\sqrt{\frac{T}{M}}X_2)\ldots \right
.  \nonumber \\ 
\left .  \ldots (1+\mu \frac{T}{M} +\sigma
\sqrt{\frac{T}{M}}X_M) \right]  
\label{Mdimprod}
\ee
where $\mu$ is some deterministic ``drift'' matrix and the stochastic
matrices $X_i$  
belong to identical independent random matrix ensembles.
In particular we will be studying the eigenvalue distribution of
$Y(T)$:
\eq
\rho_T(z)=\frac{1}{N}\cor{\tr \dl^{(2)} \left(z-Y(T)\right) }
\eqx
where the average is taken over the stochastic $N$ by $N$ matrices $X_i$ 
appearing in the definition of $Y(T)$. 

Contrary to several standard multidimensional extensions of Brownian walks,
we concentrate here on studying %the spectral flow of the operator $Y$, 
%i.e. we study 
how the full spectrum of operator $Y$ evolves with time $T$.

Let us note some key features of the operator $Y$.
First, since the matrices $X_i$ in general do not commute, 
we are dealing with a `path ordered product'.
Second, even if the matrices $X_i$ are hermitian, 
their product is not, i.e. the spectrum in general disperses into the 
complex plane, showing - as pointed out in this paper - some rather
unusual feature described as a "topological phase transition". 
Namely the support of the eigenvalue distribution changes from a
simply-connected two-dimensional island to a two-dimensional island with a
hole.

We note that the time evolution of some initial vector $|0>$
under $Y$ forms a very general setup for several multivariate 
stochastic evolutions of the type 
\be
\ket{\tau}=Y(\tau)\ket{0},
\ee
Analytical results for such processes are however scarce. 
In a recent study, Jackson et al.~\cite{ANDY} solved the matrix-valued
evolution driven by Gaussian noise of the 
type considered above in the case of 2 by 2 hermitian matrices. 
There are also some known results for the diffusion of the {\em norm}
of vector $\ket{\tau}$, since this problem can be linked to the Lyapunov 
exponents of the hermitian matrix $Y^{\dagger}Y$. We briefly comment
on this issue in section~9. 

The main result of this work is the derivation of exact (finite $M$)
and asymptotic ($M \rightarrow \infty$) 
spectral formulas for the evolution operators $Y$, in the limit where the 
dimension $N$ 
of $Y$ tends to infinity. For simplicity, we set to zero the deterministic
entries $\mu$ -- although one can easily include them using
our formalism.  
In this paper we also limit ourselves to ensembles of the Gaussian
type. Generalizations to non-Gaussian measures (involving finite and
infinite spectral supports) will be investigated elsewhere.
We discuss here two cases depending on  whether the $X_i$'s 
are (i) hermitian Gaussian matrices (GUE) or (ii) complex (arbitrary) 
Gaussian matrices (GCE).
Similar constructions could be used for real and quaternionic matrices.

The main motivation for this work is to formulate a new 
and relatively simple theoretical framework for studying spectra of the 
diffusive-like evolution operators with their further applications to
various branches of the complex  systems
analysis in physics, mathematics and in interdisciplinary 
science. 

In particular, the formalism outlined here could be used as a starting point 
for matrix versions of super- or sub-diffusions, generated by
{\it e.g.} matrix L\'evy analogs of the stable power-tailed distributions. 
It will be used also for the study of spectral statistics 
of infinite products of unitary operators ({\it e.g.} Polyakov
lines)~\cite{WIECZETAL}. 

Most of the derivations presented in this paper will be based on
diagrammatic methods. In order to make the paper self-contained, in
the first two sections we briefly present the formalism, first for
hermitian RMM (section~2) and then in section~3, we recall
the extension of diagrammatic methods to the non-hermitian
case~\cite{NONHERM,OTHERNONHERM},
by using complex Gaussian non-hermitian random matrix model as an example.
In section 4 we move to the main part of our paper and introduce our
method for the treatment of products of random matrices by performing
calculations for a product of two matrices. 
Then (section~5) we move toward the products of $M$ complex matrices, and we 
perform the limit $M\rightarrow \infty$. 
We present the final formulas for the spectral density and the equations 
defining the boundary of the two-dimensional support of the eigenvalues.
In sections~6 and 7 we consider the case of
hermitian matrices which, surprisingly, is more difficult 
than the previous one.
 We analyze first (section~6) the product of two matrices.
This is interesting on its own, since it exhibits quite
singular behavior  -- for evolution times smaller than a certain
critical value, the {\it a priori} complex eigenvalues are "frozen"
to  the real axis. 
In section~7  we move toward the general case. For
products of matrices involving more than
two hermitian matrices, the eigenvalues always form two-dimensional islands. 
We solve the spectral problem in the large $M$ limit. %, 
%providing the explicit formulae for the spectral functions and equations for 
%the boundaries shaping the support of eigenvalues. 
In section~8 we show the numerical analysis, confirming our analytical 
predictions and the existence of ``topological phase transitions'', 
corresponding to structural changes of the spectrum. 

Finally, in section~9  we briefly discuss 
the problems of stability of such products and the relation 
of our results to the standard Lyapunov exponents.
Section~10 concludes the paper.

\section{Hermitian Random Ensembles}

A key problem in random matrix theories is to find the
distribution of eigenvalues $\lambda_i $,
in the large $N$ (size of the matrix $H$)
limit, i.e
\be
\rho(\lambda)= \frac{1}{N} \left\langle \sum _{i=1}^N \delta (\lambda -
\lambda_i)
\right\rangle
\label{spect-h}
\ee
where the averaging $\cor{\ldots}$
is done over the ensemble of $N \times N$ random hermitian
matrices
generated with probability
\be
P(H) \propto e^{-N {\rm Tr} V(H)} .
\label{probab}
\ee
The eigenvalues of course lie on the real axis.
By introducing the resolvent (Green's function)
\be
G(z)=\frac{1}{N} \left\langle {\rm Tr}\, \frac{1}{{\bf z}-H}\right\rangle \,.
\label{green}
\ee
with ${\bf z}=z {\bf 1}_N$ and by using the standard relation
\be
\frac{1}{\lambda \pm i \epsilon} = {\rm  P.V.} \frac{1}{\lambda}
\mp i\pi \delta(\lambda) \,.
\label{dist}
\ee
the spectral function (\ref{spect-h}) can be derived from the 
discontinuities of the Green's function (\ref{green})
\eqn
 \frac{1}{2\pi i} \,\lim_{\epsilon \rightarrow 0} \left(
G(\lambda -i\epsilon)-G(\lambda+i\epsilon) \right)&=& \frac{1}{N}
\left\langle {\rm Tr}\, \delta(\lambda-H)\right\rangle \nonumber \\
&=& \frac{1}{N} \left\langle \sum _{i=1}^N \delta (\lambda -
\lambda_i)
\right\rangle= \rho(\lambda) \,.
\label{recon}
\eqnx

There are several ways of calculating Green's functions for 
HRMM~\cite{GENERALHER,ZINNJUSTIN,GUHR}.
We will follow  the diagrammatic approach introduced by \cite{BZ}.
A starting point of the approach is the expression allowing for the
reconstruction of the Green's function from all the moments
$\left \langle {\rm Tr} H^n \right \rangle$,
\be
G(z) &=&
        \frac{1}{N} \left\langle {\rm Tr}\,\frac{1}{{\bf z}\!-\!H}\right\rangle
=\frac{1}{N}\!\left\langle {\rm Tr}\!\left[  \frac{1}{{\bf z}} +\!%
\frac{1}{{\bf z}} H \frac{1}{{\bf z}} +\!\frac{1}{{\bf z}} H 
\frac{1}{{\bf z}} H \frac{1}{{\bf z}}
 +\!\cdots \right] \right\rangle  \nonumber \\
&=& \frac{1}{N} \sum_n  \frac{1}{z^{n+1}} \left\langle {\rm Tr} H^n
        \right\rangle 
\label{e.series}         
\ee
The reason why the above procedure works correctly for {\em hermitian}
matrix models is the fact that the Green's function is guaranteed to
be {\em holomorphic} in the whole complex plane except at most on one or
more 1-dimensional intervals.
We will use the diagrammatic method to evaluate efficiently the sum of
the moments on the right hand side. We will now restrict ourselves to the 
well known case of a random hermitian ensemble with Gaussian distribution.

The first step is to introduce a generating function with a
 matrix-valued
 source $J$:
\be
Z(J)= \int   dH
 e^{-\frac{N}{2} ({\rm
Tr} H^2) + {\rm Tr} H \cdot  J} \,.
\label{partition}
\ee
where we integrate over all $N^2$ elements of the matrix $H$.
All the moments follow directly from $Z(J)$ through the relation
\be
\cor{ {\rm Tr} H^n} = \frac{1}{Z(0)} {\rm Tr}\left(
\frac{\partial }{\partial J} \right)^n %\left.
Z(J)\Big|_{J=0}
\ee
and are straightforward to calculate, since in the Gaussian  case
the partition function (\ref{partition}) 
reads $Z(J)= \exp {\frac{1}{2N} {\rm Tr} J^2}.$
Accordingly, the lowest nonzero expectation value is
\be
\langle H^a_b H^c_d \rangle 
 =\frac{\partial^2 Z(J)}{\partial J^b_a \partial J^d_c}|_{J=0}
=\!\frac1{N} \frac{\partial J^c_d  Z(J)}{\partial J^b_a}|_{J=0}
=\!\frac{1}{N} \delta^c_b \delta ^a_d \,.
\label{prop}
\ee
and the next non-vanishing expectation value reads
\be
\langle H^a_b H^c_d H^e_f H^g_h \rangle =
\frac{1}{N^2}\left( \delta^c_b \delta^a_d \delta^g_f \delta^e_h +
\delta^a_h \delta^b_g \delta^c_f \delta^d_e +
\delta^a_f \delta^b_e \delta^c_h \delta^d_g \right)
\label{cum4}
\ee
The key idea in the diagrammatic approach is to associate to the
expressions for the moments, like the above, a graphical
representation following from a simple set of rules.
The power of the approach is that it enables to perform a resummation
of the whole power series (\ref{e.series}) through the identification
of the structure of the relevant graphs.

\begin{figure}[t]
%%%%%%%%%%%%% 1st graph %%%%%%%%%%%%%%%%%%%%%%
\centerline{\epsfysize=15mm \epsfbox{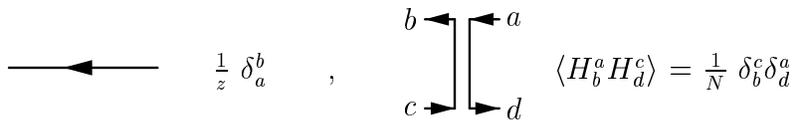}}
%%%%%%%%%%%%% 1st graph %%%%%%%%%%%%%%%%%%%%%%
\caption{Large $N$ ``Feynman" rules for ``quark'' and ``gluon" propagators.}
\label{fig.rules}
\end{figure}

\begin{figure}[t]
%%%%%%%%%%%%% 2nd graph %%%%%%%%%%%%%%%%%%%%%%
\centerline{\epsfxsize=115mm \epsfbox{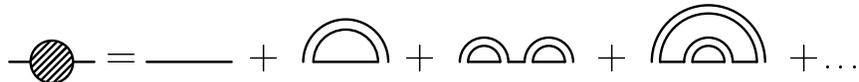}}
%%%%%%%%%%%%% 2nd graph %%%%%%%%%%%%%%%%%%%%%%
\caption{Diagrammatic expansion of the Green's function up to the $O(H^4)$ 
terms.}
\label{fig.rainbow}
\end{figure}

We depict the ``Feynman'' rules in Fig.~\ref{fig.rules}, similar to 
the standard large $N$ diagrammatics for QCD~\cite{THOOFT}. 
The  $1/{\bf z}= 1/z \delta_{a}^{b}$ in (\ref{e.series}) 
is represented by a horizontal straight line.
The propagator (\ref{prop}) 
%\be
%\langle  H^a_b H^c_d \rangle 
% = \frac{1}{N} \delta^c_b \delta^a_d  
%\ee
is depicted as a double line. 

The diagrammatic expansion of the Green's function is visualized
in Fig.~\ref{fig.rainbow}, where one connects the vertices with the
double line propagators in all possible ways.
Each ``propagator'' brings a factor of $1/N$, and each loop a
factor of $\delta^a_a=N$. 
{}From the  three terms, 
corresponding to (\ref{cum4}) contributing to $\cor{\tr H^4}$ only
the first two are presented in ~Fig.~\ref{fig.rainbow} (the third and
the fourth diagram). The diagram corresponding to the third term in
(\ref{cum4}) 
represents  a  non-planar contribution which 
is suppressed as $1/N^2$ and hence vanishes when $N \to \infty$.
In general, only planar graphs survive the large $N$ limit.

The resummation of (\ref{e.series}) is done by 
introducing the self-energy $\Sigma$ comprising the sum of all
one-particle  irreducible graphs (rainbow-like). Then the Green's
function reads 
\eq
G(z)= \frac{1}{z-\Sigma(z)}\,.
\label{e2}
\eqx

\begin{figure}[t]
%%%%%%%%%%%%% 3rd graph %%%%%%%%%%%%%%%%%%%%%%
\centerline{\epsfxsize=40mm \epsfbox{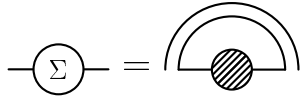}}
%%%%%%%%%%%%% 3rd graph %%%%%%%%%%%%%%%%%%%%%%
\caption{Schwinger-Dyson  equation for rainbow diagrams.}
\label{fig.pastur}
\end{figure}

In the large $N$ limit the equation for the self energy $\Sigma$,
follows from
resumming the rainbow-like diagrams of Fig.~\ref{fig.rainbow}. The
resulting equation (``Schwinger-Dyson'' equation
of Fig.~\ref{fig.pastur}) encodes pictorially the structure of these
graphs and reads
\eq
\Sigma=\frac{1}{N} {\rm Tr} G {\bf 1}=\frac{N}{N}G =G .
\label{SD}
\eqx
Equations (\ref{e2}) and (\ref{SD}) give immediately
$G(z-G)=1$ which can be solved to yield 
\be
G_{\mp}(z)=\frac{1}{2}(z\mp \sqrt{z^2-4})
\label{semi1}
\ee
Only the $G_-$ is a normalizable solution, with the proper
asymptotic behavior $G(z) \rightarrow 1/z$ in the 
$z\rightarrow \infty$ limit. From  the discontinuity (cut),
using (\ref{recon}), we recover  
Wigner's semicircle~\cite{Wigsemi}
for the distribution of the eigenvalues for hermitian random matrices
\be
\rho(\lambda)=\frac{1}{2\pi}\sqrt{4-\lambda^2}.
\ee

\section{Non-hermitian Random Ensembles}

The main difficulty in the treatment of non-hermitian random matrix
models is the fact that now the eigenvalues accumulate in {\em
two-dimensional} domains in the complex plane and 
the Green's function is no longer
holomorphic. Therefore the power series expansion (\ref{e.series}) no
longer captures the full information about the Green's function.
In particular the eigenvalue distribution is related to the
non-analytic (non-holomorphic behaviour of the Green's function:
\eq
\label{e.dzbar}
\f{1}{\pi}\partial_{\zb} G(z)=\rho(z) \ .
\eqx

This phenomenon can be easily seen even in the the simplest
non-hermitian ensemble --- the Ginibre-Girko one \cite{GinGir,GIRKO},
with non-hermitian matrices $X$, and measure
\be
P(X)= e^{-N {\rm Tr} X X^{\dagger}} \,.
\ee
It is easy to verify that all moments vanish $\cor{\tr X^n}=0$, for $n>0$
so the expansion (\ref{e.series}) gives the Green's function to be
$G(z)=1/z$ (diagrammatically this follows from the fact that the
propagator $<{X^a_b X^c_d}>$ vanishes and hence the self-energy
$\Sigma=0$). The true answer is, however, different. Only for $|z|>1$
one has indeed $G(z)=1/z$. For $|z|<1$ the Green's function is
nonholomorphic and equals $G(z)=\zb$. 

In spite of the above difficulties a diagrammatic approach can be used
for recovering the full information including the eigenvalue density
$\rho(z)$. The first step is to define a {\em regularized} Green's function
from which one can extract $\rho(z)$. This has been done by 
exploiting the analogy to two-dimensional electrostatics
\cite{HAAKE,GIRKO,SOMMERS}. The resulting regularized Green's function is
however difficult to calculate. The second step is to give an
equivalent linearized form which can form a basis for a diagrammatic
calculation~\cite{NONHERM}. We will now briefly review the above
developments.

Let us define the ``electrostatic potential''
\be
F=\frac{1}{N} {\rm Tr } \ln [({\bf z}-\MM)(\bar{{\bf z}}-\MM^{\dagger}) +
        \epsilon^2] \,.
\label{els}
\ee
Then
\eqn
\lim_{\epsilon \rightarrow 0} \frac{\partial^2 F(z,\bar{z})}%
        {\partial z\partial \bar{z}}
% \frac{\partial}{\partial \bar{z}} F(z,\bar{z}, \epsilon)
&=&\lim_{\epsilon \rightarrow 0}
\frac{1}{N}\left\langle\!{\rm Tr} \frac{\epsilon^2}{(|{\bf z}\!-\!\MM|^2
+\!\epsilon^2)^2}\!\right\rangle
\nonumber \\
 &=&
%\frac{\pi}{N} \left\langle {\rm Tr}\ \delta^{(2)}(z\!-\!\MM)
%\right\rangle=
    \frac{\pi}{N} \left\langle \sum_i \delta^{(2)}(z\!-\!\lambda_i)
        \right\rangle
\equiv
\pi \rho(x,y)
\label{spec}
\eqnx
represents Gauss law,
where $z=x+iy$. The last equality follows from 
the representation of the complex Dirac delta
\be
\pi \delta^{(2)}(z-\lm_i)=\lim_{\epsilon \rightarrow 0} 
\frac{\epsilon^2}{(\epsilon^2 +|z-\lm_i|^2)^2}
\ee
In the spirit of the electrostatic analogy we can define
the Green's function $G(z,\bar{z})$, as an ``electric field''
\be
G%(z, \bar{z})\!
\equiv\!
 \frac{\partial F}{\partial z}\!=\!
\frac{1}{N}\!\lim_{\epsilon \rightarrow 0}\!
\left\langle\!{\rm Tr}\!
\frac{\bar{{\bf z}}-X^{\dagger}}{(\bar{{\bf z}}\!-\!X^{\dagger})({\bf z}\!-\!X)
+\!\epsilon^2)}\!\right\rangle .\!
\label{GG}
\ee
Then Gauss law reads 
\be
\partial_{\zb}G=\pi \rho(x,y)
\ee
in agreement with (\ref{e.dzbar}).

As mentioned above, instead of working {\it ab initio} with the object
(\ref{GG}), and in view of applying diagrammatic %and free-random variable
methods it is much more convenient to proceed differently --  as we will
now discuss.

Following~\cite{NONHERM} we define the matrix-valued resolvent
through
\be
\hat{{\cal {G}}}
&=&\frac{1}{N} \left\langle \rm{Tr_{B2}}
\setlength\arraycolsep{0pt}
\arr{{\bf z}-X}{i  \epsilon}{i  \epsilon}{{\bf \zb}
-X^{\dagger}}^{-1}\right\rangle = \nonumber\\
&=&
\f{1}{N} \left\langle {\rm Tr_{B2}} 
\arr{A}{B}{C}{D} \right\rangle \equiv
\setlength\arraycolsep{3pt}
\arr{{\GG}_{11}}{{\GG}_{1\overline{1}}}{{\GG}_{\overline{1}1}}
{{\GG}_{\overline{1}\overline{1}}}
\label{19}
\ee
with 
\be 
A&=&\frac{{\bf \zb} -X^{\dagger}}{({\bf
          \zb} -X^\dagger)({\bf z}-X)+\epsilon^2} \nonumber \\
B&=&\frac{-i\epsilon }{({\bf
          z} -X)({\bf \zb}-X^{\dagger})+\epsilon^2} \nonumber \\
C&=&\frac{ -i\epsilon}{({\bf
          \zb} -X^\dagger)({\bf z}-X)+\epsilon^2} \nonumber \\
D&=&\frac{ {\bf z} -X}{({\bf
          z} -X)({\bf \zb}-X^{\dagger})+\epsilon^2} 
\ee
and 
where we introduced the `block trace'  defined as
\be
{\rm Tr_{B2}}
\setlength\arraycolsep{3pt}
\arr{A}{B}{C}{D}_{2N \times 2N} \hspace*{-3mm}\equiv
\setlength\arraycolsep{3pt}
\arr{
{\rm Tr}\ A}{{\rm Tr}\ B}{{\rm Tr}\ C}{{\rm Tr}\ D}_{2 \times 2}
\hspace*{-5mm}\,.
\label{block2}
\ee
Then, by definition, the upper-right component $\GG_{11}$, is equal to
the Green's function (\ref{GG}).

The block approach has several advantages. First of all it is {\em
linear} in the random matrices $X$ allowing for a simple
diagrammatic calculational procedure.
Let us define $2N$ by $2N$ matrices
\be
\label{defzg}
{\mbox{\boldmath $\cal Z $}}=
\arr{\bf{z}}{i \epsilon {\bf 1}}{i 
 \epsilon {\bf 1}}{\bar{{\bf z}}} \quad, \quad
\HH=\arr{X}{0}{0}{X^{\dagger}} \,.
\ee
Then the generalized Green's function is given formally by the same
definition as the usual Green's function $G$,
\be
\GG=\frac{1}{N} \left\langle {\rm Tr_{B2}} 
\frac{1}{{\mbox {\boldmath $\cal Z$   } }-\HH}\right\rangle
        \,.
\label{concise}
\ee
What is more important, also in this case the Green's function is completely
determined by the knowledge of all  matrix-valued moments
\be
\left\langle {\rm Tr_{B2}}\,\,\, {\mbox{\boldmath $ \ZZ$}}^{-1} \HH 
{\mbox{\boldmath $ \ZZ$}}^{-1}
 \HH \ldots {\mbox{\boldmath $\ZZ$}}^{-1}
        \right\rangle \,.
\label{genmom}
\ee
This last observation allows for a diagrammatic interpretation. The
Feynman rules are analogous to the hermitian ones, only now one has to
keep track of the block structure of the matrices, e.g. single
straight lines will now be associated with a {\em matrix} factor
$\ZZ^{-1}$. We will now demonstrate the above procedure by solving
diagrammatically the complex Gaussian Random Matrix Model.

\subsubsection*{Example: the Girko-Ginibre ensemble}

The ensemble is defined by the measure
\be
P(X) \propto e^{-N {\rm Tr} X X^{\dagger}} \,.
\ee
In this case the double line propagators are
\be
\langle X^a_b X^c_d \rangle  &=& \langle {X^\dagger}^a_b
{X^\dagger}^c_d \rangle = 0  \nonumber\\ 
\la X^a_b {X^\dagger}^c_d \rangle &=& \la {X^\dagger}^a_b X^c_d \rangle =
        \frac{1}{N}\delta^a_d \delta_c^b \,.
\label{comprop}
\ee
As previously, we introduce the self-energy $\tilde{\Sigma}$, (but
which is now matrix-valued), in terms of which we get a two by two
matrix expression  
\be
\GG=({\ZZ}-\tilde{\Sigma})^{-1} .
\label{gin1}
\ee
where the 2 by 2 matrix $\ZZ$ reads
\be 
{\cal Z}= \arr{z}{i\epsilon}{i\epsilon}{\zb}
\ee
The resummation of the rainbow diagrams for $\tilde{\Sigma}$ is more
subtle, but follows easily from the structure of the propagators
(\ref{comprop}). 
%Instead of the hermitian equation $\Sigma=G\cdot 0=0$ (due to the
%vanishing of the first propagator in (\ref{comprop})), 
The analogue of (\ref{SD}) is now:
\eqn
\tilde{\Sigma} &\equiv&
\setlength\arraycolsep{3pt}
\arr{\Sigma_{11}}{\Sigma_{1\bar{1}}}{\Sigma_{\bar{1}1}}%
        {\Sigma_{\bar{1}{\bar{1}}}}=
%\setlength\arraycolsep{1pt}
%\arr{\GG_{11}\cdot
%0}{\GG_{1\bar{1}}}{\GG_{\bar{1}1}}{\GG_{\bar{1}\bar{1}}\cdot 0}=
%\nonumber\\
%
\setlength\arraycolsep{1pt}
\arr{0}{\GG_{1\bar{1}}}{\GG_{\bar{1}1}}{0}
\,.
\label{gin2}
\eqnx
The two by two  matrix equations~(\ref{gin1}-\ref{gin2}) completely
determine the problem of finding the eigenvalue distribution for
the Girko-Ginibre ensemble. Inserting (\ref{gin2}) into (\ref{gin1}) we get:
\eq\setlength\arraycolsep{3pt}
\arr{{\GG}_{11}}{{\GG}_{1\overline{1}}}{{\GG}_{\overline{1}1}}
{{\GG}_{\overline{1}\overline{1}}}=
\f{1}{|z|^2-\GG_{1\bar{1}}\GG_{\bar{1}1}} \cdot
\arr{\zb}{{\GG}_{1\overline{1}}}{{\GG}_{\overline{1}1}}{z} .
\label{e.sdgin}
\eqx
Note that at this moment we can safely put to zero the regulators $\epsilon$.
Looking at the off-diagonal equation
\be
\GG_{1\bar{1}}=\frac{\GG_{1\bar{1}}}{|z|^2-\GG_{1\bar{1}}\GG_{\bar{1}1}}
\ee
we see that there are two solutions: one
with $\GG_{1\bar{1}}=0$, and another with $\GG_{1\bar{1}} \neq 0$.
The first one leads to a holomorphic Green's function, and a
straightforward calculation gives
\be
G(z)=\frac{1}{z} \,.
\ee
The second one is nonholomorhic, imposing the condition
\be
|z|^2 -b^2=1
\label{circle}
\ee
where we denoted $\GG_{1\bar{1}}\GG_{\bar{1}1}\equiv b^2$, hence
\be
G(z, \bar{z}) =\bar{z} 
\ee
which leads, via the Gauss law, to 
\be
\rho(x,y)=\frac{1}{\pi}
\frac{\partial}{\partial \bar{z}} G(z,\bar{z}) =\frac{1}{\pi} \,.
\ee
Both solutions match at the boundary $b^2=0$ , which in this case reads
$z \bar{z}=1$.  In such a simple way we recovered
the results of Ginibre and Girko for the complex non-hermitian ensemble.
The eigenvalues are uniformly distributed on the unit disk $|z|^2<1$.

This example illustrates more general properties of the matrix valued
generalized Green's function. Each component of the matrix carries important
information about the stochastic properties of the system.
There are always two solutions for $\GG_{11}$, one holomorphic, another
non-holomorphic. The second one leads, via Gauss law, to the eigenvalue
distribution. The shape of the ``coastline'' bordering the ``sea'' of
complex eigenvalues is determined by the
matching conditions for the two solutions,
i.e. it is determined by imposing on the non-holomorphic solution 
for $b^2$ the equation $b^2=0$.
%$\GG_{1\bar{1}}\GG_{\bar{1}1}=0$.
For other important features of the explained above diagrammatic
method and more complex examples see \protect\cite{NONHERM,USEVECT}.

\section{Product of Two Complex Matrices}

We will now move toward the main goal of this paper, namely the
evaluation of the spectral properties of the random matrix products of
the form (\ref{Mdimprod}) -- with zero drift and normalized variance:
\be
Y(T)= \lim_{M \to \infty} Y_M\equiv \lim_{M \to \infty} 
\left [ (1 + \sqrt{\frac{T}{M}}X_1) 
(1 +\sqrt{\frac{T}{M}}X_2)\ldots \right.  \nonumber \\ 
\left .  \ldots (1+ \sqrt{\frac{T}{M}}X_M) \right] 
\label{e.master}
\ee
In principle the methods of section 3 could be used to study such
random matrices, however the nonlinear product structure would make
the resulting diagrammatic rules exceedingly difficult to
control. Fortunately, for the purposes of calculating the Green's
function, and hence the eigenvalue density, the problem may be {\em
linearized} at the cost of increasing the size of the matrices. In
this section we will show how the method works for $Y_2$ -- the
product of two matrices, while in section 5 we will consider the
general case including the limit $M \to \infty$.  
We therefore have to calculate  
\be
Y_2= 
\left(1+\sqrt{\frac{\tau}{2}}X_1\right) \left(
1+\sqrt{\frac{\tau}{2}}X_2\right) 
\equiv A_1 A_2
\label{case2}
\ee
where $X_1$ and $X_2$ belong to identical independent Gaussian Complex
Ensembles (iiGCE=Girko-Ginibre). Since the matrices are non-hermitian
we have to use the formalism of section 3 with a conjugate
`antiholomorphic' copy.  
% An essential complication, comparing to the case of 
% the Ginibre-Girko ensemble, is  that 
% now all the 
% generalized moments include same powers of  $A_i\equiv{\bf{1}} +
% \sqrt{\frac{\tau}{2}}X_i$.
>From now on in the rest of the paper we set the regulator $\epsilon$
to zero. 
In order to linearize the product structure we use a trick:
we introduce a further doubling, thus 
constructing  an auxiliary $4N$ by $4N$ Green's function
\be
\GG(w) &=& \left\langle\left[ \left( 
\begin{array}{cccc} {\bf w}&0&0&0 \\ 0&{\bf w}&0&0 \\ 0&0&\bar{{\bf w}}&0
 \\ 0&0&0&\bar{{\bf w}}
\end{array}
\right)  
- \left( 
\begin{array}{cccc} 0&A_1&0&0 \\ A_2&0&0&0 \\ 0&0&0&A_2^{\dagger} \\
 0&0&A_1^{\dagger}&0
\end{array}\right)  
\right]^{-1}_{4N \times 4N}\right\rangle \nonumber \\
&=& \left\langle\left[ \left( 
\begin{array}{cccc} {\bf w}&-{\bf 1}&0&0 \\ -{\bf 1} & {\bf w}&0&0 \\ 
0&0&\bar{{\bf w}}&-{\bf 1} \\ 0&0&-{\bf 1}&\bar{{\bf w}}
\end{array}
\right)  
- \sqrt{\frac{\tau}{2}} \left( 
\begin{array}{cccc} 0&X_1&0&0 \\ X_2&0&0&0 \\ 0&0&0&X_2^{\dagger} \\
 0&0&X_1^{\dagger}&0
\end{array}\right)  
\right]^{-1}_{4N \times 4N}\right\rangle \nonumber \\
 &\equiv&  \left\langle\left[{\mbox{\boldmath  $\cal W$}}-
\sqrt{\tau/2}{\cal{X}}\right]^{-1}\right\rangle 
\label{case2green}
\ee
%Here $A_i \equiv 1+\sqrt{\tau/2}X_i$.
In the above we first separated the ``deterministic'' part from the
``random one'' (second line) and in the last equality 
we introduced the notation similar to the hermitian
and non-hermitian  cases analyzed in the previous sections.

In analogy to (\ref{block2}), 
we define now a block-trace operation $\tr_{B4}$, where
we trace each $N$ by $N$ block of the $4N$ by $4N$ matrix $\GG (w) $
separately. In such a way, we obtain 
a 4 by 4 auxiliary Green's function 
\be
g(w) \equiv \left(  
\begin{array}{cccc} g_{11}&g_{12}&g_{1\bar{1}}&g_{1\bar{2}} \\ 
g_{21}&g_{22}&g_{2\bar{1}}&g_{2\bar{2}} \\ 
g_{\bar{1}1}&g_{\bar{1}2}&g_{\bar{1}\bar{1}}&g_{\bar{1}\bar{2}} \\ 
g_{\bar{2}1}&g_{\bar{2}2}&g_{\bar{2}\bar{1}}&g_{\bar{2}\bar{2}} \\
\end{array}
\right)_{4 \times 4} = \frac{1}{N} \tr_{B4}\GG(w)
\label{case2green4}
\ee
The advantage of this function lies in the fact, that the 
eigenvalues of the product $A_1 A_2$ coincide with the squares of the
eigenvalues of the $2N\times 2N$ block matrix
\eq
\label{e.blocktwo}
\arr{0}{A_1}{A_2}{0}
\eqx
This is due to the off-diagonal structure of the sub-blocks in
(\ref{e.blocktwo}). For a general discussion for an arbitrary product
see section 5. 

% moments of 
% the function
% $g(w)/w$, where $w^2=z$, are identical to the moments of the 
% desired generalized Green's function (in the sense of section 3) for
% the product $A_1 A_2$.  
% This is due to the off-diagonal structure 
% of the sub-blocks in (\ref{case2green}). 
% Only even moments in ${\cal{X}}$ contribute to the trace, 
% and the odd, by definition, have zeroes on the diagonal.   

As before we define a 4 by
4 matrix of self-energies $\tilde{\Sigma}_{ij}$
\be
g(w)= ({\cal{W}}-\tilde{\Sigma})^{-1}
\ee
where ${\cal W}$ is the result of block-tracing 'bold' 
$\mbox{\boldmath $\cal W$} $.

Similarly as in the Girko-Ginibre case, we can now diagrammatically 
analyze the content of the matrix $\tilde{\Sigma}$. 
We see that only the ``double line'' propagators 
$\cor{X_1 X_1^{\dagger}}_{X_1}$ and  $\cor{X_2 X_2^{\dagger}}_{X_2}$
are different from zero.  
Therefore from all of the 16 elements of matrix $\tilde{\Sigma}$ only four 
are nonzero. Moreover, due  to  identical measures
of $X_1$ and $X_2$ variables, we get
\be
\Sigma_{1\bar{1}}&=&\Sigma_{2\bar{2}}=\alpha g_{\bar{1}1}=\alpha g_{\bar{2}2} \nonumber \\
\Sigma_{\bar{1}1}&=&\Sigma_{\bar{2}2}=\alpha g_{1\bar{1}}=\alpha g_{2\bar{2}} 
\label{sigmael} 
\ee 
where the factor $\alpha=\tau/2$ comes from the propagator in the 
corresponding 
Schwinger-Dyson  equation of the type discussed in the previous chapter.
 
Filling  the matrix $\tilde{\Sigma}$  with entries (\ref{sigmael})
we arrive at the 4 by 4 matrix  equation for the elements of the Green's
function:
\eqn
\left(  
\begin{array}{cccc} g_{11}&g_{12}&g_{1\bar{1}}&g_{1\bar{2}} \\ 
g_{21}&g_{22}&g_{2\bar{1}}&g_{2\bar{2}} \\ 
g_{\bar{1}1}&g_{\bar{1}2}&g_{\bar{1}\bar{1}}&g_{\bar{1}\bar{2}} \\ 
g_{\bar{2}1}&g_{\bar{2}2}&g_{\bar{2}\bar{1}}&g_{\bar{2}\bar{2}} \\
\end{array}
\right)_{4 \times 4}=
&& \left[  \left( 
\begin{array}{cccc} {\bf w}&-{\bf 1}&0&0 \\ -{\bf 1} & {\bf w}&0&0 \\ 
0&0&\bar{{\bf w}}&-{\bf 1} \\ 0&0&-{\bf 1}&\bar{{\bf w}}
\end{array}
\right) \right. - \nonumber\\
&&-\left. \left(  
\begin{array}{cccc} 
0 & 0 & \al g_{\bar{1}1} & 0 \\ 
0 & 0 & 0 & \al g_{\bar{1}1} \\ 
\al g_{1\bar{1}} & 0 & 0 & 0 \\ 
0 & \al g_{1\bar{1}} & 0 & 0 \\
\end{array}
\right)
\right]^{-1}
\eqnx

As in the Ginibre case, we solve it for $g_{\bar{1}1}$ and $g_{1\bar{1}}$.
Then, using the solution, we calculate $g_{11}$. 
Coming back to original variables with $w^2=z$ we set $G(z,\bar{z})=g_{11}(w)$.
The eigenvalue distribution then follows from 
\be
\label{e.rhotwo}
\rho(x,y)=\frac{1}{\pi}\f{w}{z}\partial_{\bar{z}}G(z,\bar{z})
\ee
This is a consequence of the relation between the eigenvalues of the block
matrix (\ref{e.blocktwo}) and the eigenvalues of the product
$A_1A_2$. A general derivation is outlined in the following section.

A little algebra (basically an inversion of the 4 by 4 matrix) shows, that, 
as before, we get two solutions for 
$b^2=g_{1\bar{1}}g_{\bar{1}1}$$=g_{2\bar{2}}g_{\bar{2}2}
$:  (i)  the holomorphic one 
(corresponding to $b^2=0$, i.e. all $\Sigma_{ij}$  equal to zero); 
(ii) the non-holomorphic one,
given by the equation  
\be
\frac{\tau}{2}\left(|w|^2-\alpha^2 b^2+1\right)=
(\alpha^2 b^2-1+w^2)(\alpha^2 b^2-1+\bar{w}^2)-
\alpha^2 b^2(w+\bar{w})^2
\label{gap2}
 \ee
The boundary of the eigenvalue support is given, as before, 
by the above equation with $b^2$ set to 0.
It is therefore a fourth order conchoid-like planar curve,  
which in polar coordinates is given by 
\be
\frac{\tau}{2} (1+r)=r^2+1-2r\cos \phi
\label{conchoid}
\ee

\begin{figure}[t]
%%%%%%%%%%%%% conchoid %%%%%%%%%%%%%%%%%%%%%%
\centerline{\epsfysize=55mm \epsfbox{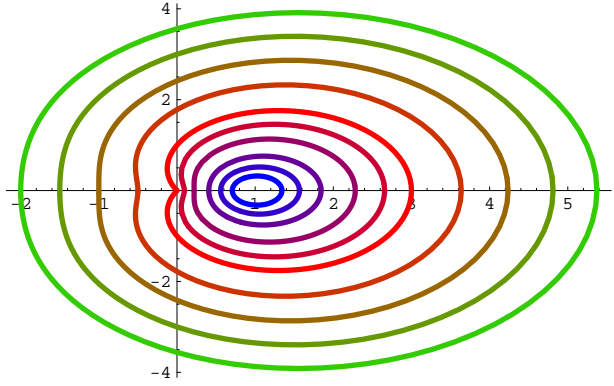}}
%%%%%%%%%%%%% 1st graph %%%%%%%%%%%%%%%%%%%%%%
\caption{Evolution of the contour of the non-holomorphic domain 
in the eigenvalue spectrum of the product of two Ginibre-Girko
ensembles
as a function of evolution times $\tau=0.1,0.25,0.5,1,1.5,2,3,4,5,6$,
rising from inside to outside.}
\label{fig.konch}
\end{figure}

On Fig.~\ref{fig.konch}  we show the shape of the nonholomorphic
island 
for several sample ``evolution times'' $\tau$.   

The Green's function for the product follows from
\eq
g_{11}(w)=\f{w(\bar{w}^2-1)-\alpha^2 b^2 \bar{w}}{det}
\eqx
where
\eq
det=(\alpha^2 b^2-1+w^2)(\alpha^2 b^2-1+\bar{w}^2)- \alpha^2 b^2(w+\bar{w})^2
\eqx

\section{Product of Arbitrary Number of Complex Matrices}

We consider now the product of an arbitrary number of matrices.
To see the pattern, let us look briefly at the case of three
matrices
\be
Y_3&=& 
\left(1+\sqrt{\frac{\tau}{3}}X_1\right) \left(
1+\sqrt{\frac{\tau}{3}}X_2\right)\left(
1+\sqrt{\frac{\tau}{3}}X_3\right)  \nonumber \\
&\equiv&  A_1 A_2 A_3 
\label{case3}
\ee
where $X_1,X_2,X_2$ again belong to the Girko-Ginibre ensemble. 

Our approach again follows from a very simple {\em exact} relation
between the eigenvalues of a product of $N\times N$ matrices $A_1 A_2
A_3$ and the eigenvalues of a block matrix
\eq
\label{e.bb3}
\BB_3 = \left( 
\begin{array}{ccc} 0&A_1&0 \\0&0& A_2\\ 
A_3&0&0
\end{array}\right)_{3N \times 3N} 
\eqx
Indeed if $\{\lm_i\}$ are the eigenvalues of $A_1 A_2 A_3$ then the
eigenvalues of the block matrix are $\{ \lm_i^{1/3}, \lm_i^{1/3} \cdot
e^{\f{2\pi i}{3}}, \lm_i^{1/3} \cdot e^{\f{4\pi i}{3}} \}$. This
is an exact relation for any $N$ and follows from the relation between
the resolvents (here the matrices $A_i$ are of {\em finite size} and
{\em fixed} i.e. no averaging) 
\eq
G_{\BB_3}(w)=\f{1}{3N} \tr \f{1}{w-\BB_3} \qqqq 
G_{A_1 A_2 A_3}(z)= \f{1}{N} \tr \f{1}{z-A_1 A_2 A_3}
\eqx
namely
\eq
w  G_{\BB_3}(w) = z  G_{A_1 A_2 A_3}(w^3\equiv z)
\eqx
This is due to the cyclic structure of the block matrix (\ref{e.bb3}).
Obviously, only multiplicities of the cubic powers of $\BB_3$
contribute to the trace. 

The relation between the eigenvalues now follows from the location of
the {\em finite} number of poles of both functions.

We can thus safely calculate the eigenvalue density for the block
matrix $\BB_3$ using the diagrammatic methods and then extract the
density for the product through
\eq
\rho_{A_1A_2\ldots A_M}(z,\bar z) =
\f{1}{\pi} \partial_{\bar{z}} G_{A_1 A_2\ldots A_M}
= \f{1}{M} \f{w\bar w}{z\bar z}
\rho_{\BB_M}(w, \bar w)
\label{densref}
\eqx
where $M=3$ and $w^M=z$ and $\rho_{\BB_M}(w,\bar{w})=\f{1}{\pi}
\partial_{\bar{w}}G_{\BB}(w,\bar{w})$. 
%In terms of the auxillary Green's function we
%thus have...

To extract only the moments involving powers of the triples
$A_1A_2A_3$,   we  
construct  an auxiliary $6N$ by $6N$ Green's function, triplicating
the $A_1A_2A_3$ products by rewriting them as cyclic block matrices:
\eqn
\GG(w)= && 
\left\langle \left[ \left( 
\begin{array}{cccccc} 
{\bf w}&-{\bf 1} & 0 & 0 &0&0 \\ 
0&{\bf w} &-{\bf 1} & 0& 0& 0 \\ 
-{\bf 1} & 0 & {\bf w} & 0&0&0\\ 
0 & 0& 0&{\bf \bar{w}}&0 & -{\bf 1} \\
0&0&0&-{\bf 1} &{\bf \bar{w}}&0  \\
 0&0&0&0&-{\bf 1} & {\bf \bar{w}}
\end{array}
\right)  \right. 
\right. 
\nonumber\\
&& - \left. \left. \sqrt{{\frac{\tau}{3}}} 
\left( 
\begin{array}{cccccc} 0&X_1&0&0&0&0 \\0&0& X_2&0&0&0 \\ 
X_3&0&0&0&0&0\\ 0&0&0&0&0& X_3^{\dagger} \\ 0&0&0&X_1^{\dagger}&0&0 \\
 0&0&0&0&X_2^{\dagger}&0
\end{array}\right)  
\right]^{-1}_{6N \times 6N}\right\rangle 
\label{case3green}
\eqnx
where again  we separated the ``deterministic'' part from the ``random one''.
Introducing now the block-trace operation $\tr_{B6}$, we obtain a 6 by 6 
auxiliary  Green's function $g(w)=\tr_{B6}\GG(w)$.

% As previously  all the moments of this function are identical to 
% the moments of the wanted Green's  function for $A_1 A_2 A_3$, provided we 
% identify $w^3=z$ and we identify $G(z)=w^{-2}g(w)$. This is due to the 
% cyclic structure of the diagonal subblocks
% \be
%   \left( 
% \begin{array}{ccc} 0&X_1&0 \\0&0& X_2\\ 
% X_3&0&0
% \end{array}\right)_{3N \times 3N} 
% \label{case3greenbis}
% \ee
% Obviously, 
% only multiplicities of the cubic powers of the above block
% contribute to the trace. 

The generalization for  arbitrary $M$ is now straightforward. For 
\be
Y_M=
\left(1+\sqrt{\frac{\tau}{M}}X_1\right) \left(
1+\sqrt{\frac{\tau}{M}}X_2\right)  \dots \left(
1+\sqrt{\frac{\tau}{M}}X_M\right) 
\label{caseM}
\ee
we define an auxiliary $2MN$ by $2MN$ Green's function
of the form 
\be
\GG(w)=   \left\langle\left( 
\begin{array}{cc} {{\mbox{\boldmath $\cal
W$}}}&-\sqrt{\frac{\tau}{M}}{\cal X} \\  
-\sqrt{\frac{\tau}{M}}{\cal X}^{\dagger}& {{\mbox{\boldmath $\cal
W$}}}^{\dagger} 
\end{array}\right)^{-1}\right\rangle  
\ee
where the blocks 
\be 
{\mbox{\boldmath $\cal W$}}= \left( 
\begin{array}{ccccc} 
{\bf w}  & -{\bf 1} & 0 & \dots & 0 \\
   0     &  {\bf w} & -{\bf 1}& \dots & 0\\
  \dots & \dots & \ldots & \ldots & \ldots \\
  0& 0& \ldots & {\bf w} & -{\bf 1} \\
   -{\bf 1} & 0 & \ldots & 0 & {\bf w}
\end{array} 
 \right)_{MN \times MN}   
\ee
and 
\be 
{\cal X}=  \left( 
\begin{array}{ccccc} 
0  & X_1 & 0 & \dots & 0 \\
   0     &  0 & X_2& \dots & 0\\
  \dots & \dots & \ldots & \ldots & \ldots \\
  0& 0& \ldots & 0 & X_{M-1} \\
  X_M & 0 & \ldots & 0 & 0
\end{array} 
 \right)_{MN \times MN}   
\ee
are themselves  $NM$ by $NM$ matrices, i.e.  each of the listed elements
in {\boldmath ${\cal W}$} and in ${\cal X}$ 
is itself an $N$ by $N$ matrix, either diagonal, denoted by a bold symbol, 
or a random entry $X_i$, otherwise an $N$ by $N$ block of zeroes.  

We take now 
a  block-trace operation $\tr_{BM}$, where
we trace each $N$ by $N$ block of the $2MN$ by $2MN$ matrix $\GG (w) $
separately. In such a way, we obtain 
an  $2M$ by $2M$ auxiliary Green's function 
\be
g(w) \equiv \left(  
\begin{array}{cccccc}
 g_{11}& \ldots & g_{1M}&g_{1\bar{1}}&\ldots& g_{1\bar{M}} \\ 
\vdots &  \ddots & \vdots & \vdots & \ddots & \vdots      \\ 
 g_{M1}&\ldots& g_{MM}& g_{M\bar{1}}& \ldots & g_{M\bar{M}} \\
 g_{\bar{1}1}& \ldots  & g_{\bar{1}M} & g_{\bar{1}\bar{1}}& \ldots& 
g_{\bar{1}\bar{M}} \\ 
\vdots &  \ddots & \vdots & \vdots & \ddots & \vdots\\ 
g_{\bar{M}1}& \ldots & g_{\bar{M}M}& g_{\bar{M}\bar{1}}& \ldots &
g_{\bar{M}\bar{M}} \\
\end{array}
\right)_{2M \times 2M} = \frac{1}{N}{\rm Tr}_{BM} {\GG}(w)
\label{caseMgreenM}
\ee
%{\it $\bullet$ check if $1/N$ or $1/MN$ due to the symmetry of traces}
%By construction,   the  moments of 
%the function
%$g(w)/w^{M-1}$, where $w^M=z$, are identical to the moments of the 
%desired Green's function for the $Y_M$. 
As before we define a $2M$ by
$2M$ matrix of self-energies $\tilde{\Sigma}_{ij}$
\be
g(w)= 
\left[  
\arr{{\cal{W}}}{0}{0}{{\cal W}^{\dagger}}
 -\tilde{\Sigma}
\right]^{-1}
\ee
where ${\cal W}$ is a result of block-tracing ${\mbox{\boldmath $\cal W$}}$. 
We can now diagrammatically 
analyze the content of the matrix $\tilde{\Sigma}$. 
As previously, only ``double line'' propagators 
$\cor{X_i X_i^{\dagger}}_{X_i}$
for $i=1,\ldots, M$ are different from zero. 
Therefore from all of the $4M^2$ elements of the matrix
$\tilde{\Sigma}$ only $2M$ are different from zero. Due to the
symmetries we get
\be
\Sigma_{1\bar{1}}&=&\ldots=\Sigma_{M\bar{M}}=\alpha g_{\bar{1}1}=\alpha g_{\bar{M}M} \equiv \alpha g \nonumber \\
\Sigma_{\bar{1}1}&=&\ldots=\Sigma_{\bar{M}M}=\alpha g_{1\bar{1}}=\alpha g_{1\bar{M}} \equiv \alpha \tilde{g} 
\label{sigmaelM} 
\ee 
where the factor $\alpha=\tau/M$ comes from the propagator in the 
corresponding set of 
Schwinger-Dyson  equations as in the previous chapter.
 
Inserting  now the matrix $\tilde{\Sigma}$  with entries (\ref{sigmaelM})
we arrive at a $2M$ by $2M$ matrix  equation for the elements of the Green's
function. First, we 
solve it for $g$ and $\tilde{g}$.
Second, using the solutions, we calculate $g_{11}$. 
Third, using (\ref{densref}) 
%we get 
%recover the original function 
% $G(z,\bar{z})=\frac{g_{11}(w)}{w^{M-1}}$.
 we get an explicit equation for the spectral density
\be
\rho(x,y)=\frac{1}{\pi}\frac{1}{M} \frac{w \bar{w}}{z \bar{z}}
\partial_{\bar{w}}g_{11}(w,\bar{w})
\label{explicitdensity}
\ee
where $z=x +iy$. For arbitrary $M$, the algebra is rather involved. Luckily, both 
the block  and the cyclic structure of the main entries ${\cal W}$
and $\tilde{\Sigma}$ make taking an inverse of the $2M$ by $2M$ matrix
possible. The inverse of a cyclic matrix 
is a cyclic matrix, and its explicit form can be obtained from 
the solution of an associated recurrence relation, e.g.
using the transfer matrix techniques which we will now proceed to do.

At this moment, it is useful to introduce the notation
\be
1+|w|^2 -\alpha^2 b^2 &=& 2|w| \cosh u  \nonumber \\
|w|/w&=&\delta
\ee
The inversion needed for the Green's function reduces 
to finding explicit forms of the cyclic $M$ by $M$ matrices ${\cal
C}$, $\tilde{\cal C}$ 
defined as 
\be 
{\cal C} = [{\cal{W}}\cdot {\cal{W}}^{\dagger}-\alpha^2 b^2 {\bf
1}]^{-1} \qqqq
\tilde{\cal C} = [{\cal{W}}^\dagger\cdot {\cal{W}}-\alpha^2 b^2 {\bf
1}]^{-1}
\ee
where $b^2=g\tilde{g}$. $\tilde{\cal C}$ is just the transpose, or
equivalently, the complex conjugate of ${\cal C}$.
The elements $c_0,\ldots c_{M-1}$ of the first row of the cyclic
matrix ${\cal C}$ fulfill the recurrence relation
\eq
\label{e.rekdef}
\left(
\begin{array}{c}
 c_{k+1} \\ c_k
\end{array} 
\right) =\arr{ 2\delta \cosh u}{ -\delta^2}{ 1}{0} 
\left( \begin{array}{c} c_k \\ c_{k-1}
 \end{array} 
\right)
\eqx
with the `boundary conditions'
% \eqn
% c_M &\equiv & c_0 \\
% 2 \dl w \cosh u c_0-w c_1-\bar{w} c_{M-1} &=& 1
% \eqnx
\be
\left(
\begin{array}{c}
 c_1 \\ c_0
\end{array} 
\right) =\arr{ 2\delta \cosh u}{ -\delta^2}{ 1}{0} 
\left( \begin{array}{c} c_M \\ c_{M-1}
 \end{array} 
\right)
-\frac{1}{w} 
\left(
 \begin{array}{c} 1 \\ 0 
\end{array} 
\right)
\label{reccurrence}
\ee
The above recurrence can be easily solved using transfer matrix techniques, 
i.e. by diagonalizing the 2 by 2 transfer matrix in (\ref{e.rekdef})
and returning at the end of the calculations to the original basis. 
An explicit solution for the cyclic string $c_k$, for $k=0,1,..,.M-1$
reads
\be
\label{e.ck}
c_k=\frac{1}{w(\Lambda_+ -\Lambda_-)}\left(\frac{\Lambda_+^k}{\Lambda_+^M-1}
- \frac{\Lambda_-^k}{\Lambda_-^M-1} \right)
\ee
where $\Lambda_{\pm}=\delta e^{\pm u}$ denote the eigenvalues of the
transfer matrix. 

We may now express our key quantities for arbitrary $M$ in terms of
(\ref{e.ck}):
\be
g_{11}(w)&=& [{\cal{W}}^{\dagger} \cdot {\cal{C}}]_{11} \equiv \bar{w}
\cdot c_0 -c_{M-1}^* \nonumber \\
g&=& \alpha g [\tilde{\cal C}]_{11} \equiv \alpha g \cdot
c_0^* \nonumber \\ 
\tilde{g}&=& \alpha \tilde{g} [{\cal C}]_{11} \equiv \alpha \tilde{g}
\cdot c_0
\ee
where ${}^*$ denotes complex conjugation.
As always in the case of the complex spectra, we have two solutions:
trivial (holomorphic), corresponding to $g=\tilde{g}=0$,
and the nontrivial (non-holomorphic), corresponding to the case
$b^2\equiv g \tilde{g} \neq 0$. 

Inserting the nonholomorphic solution into $g_{11}$, 
redefining $G(z,\zb)=g_{11}(w)$ with $w^M=z$, and using
(\ref{densref}) yields the eigenvalue density for $Y_M$. 
Since $\partial_{\bar{w}}\bar{z}=
M \bar{z}/{\bar{w}}$, the result for the final spectral function
can be further simplified
\be
\rho(x,y)=\frac{1}{\pi} \frac{w}{z} \partial_{\bar{z}}G(z,\bar{z})
\label{infdensm}
\ee

%Inserting the nonholomorphic solution into $g_{11}$, 
%redefining $G(z,\zb)=G(w)/w^{M-1}$ with $w^M=z$ and taking the final 
%derivative $\pi \partial_{\zb} G(z,\zb)$ gives the spectral function.

Explicit formulas for arbitrary $M$ could be easily constructed 
e.g. by running any symbolic  program using our solution. 
The shape of the boundary is given  by the condition 
that the nonholomorphic solution meets the holomorphic one on the
complex plane.  

After solving the problem for arbitrary $M$, we can now address the 
original problem, i.e. the solution for a continuous product
of matrices, corresponding to the limit $M \rightarrow \infty$. 
This means, that we have to perform a careful limiting procedure 
in the above formulas. 
Introducing $U=Mu$, expanding in $M$, and using the polar
parametrization $z=r \exp i\phi$, we get the  solution
\be
G(z, \bar {z})= \frac{1}{\tau} \log r +\frac{1}{2} -\frac{iU}{\tau}
\f{\sin{\phi}}{\sinh U}
\label{finalcomplex}
\ee
where  $U=\sqrt{\log^2 r -\tau^2 b^2}$ with $b^2$ fulfilling 
the equation 
\be 
\cos \phi= \cosh U -\frac{\tau}{2U} \sinh U
\label{finalgap}
\ee
The set of equations (\ref{finalcomplex},\ref{finalgap}) completes 
the solution of the continuous product of large complex matrices with
Gaussian disorder. We would like to note, that due to the limiting 
procedure $M \rightarrow \infty$ differentiation of the Green's
functions has to be done with care. 
Taking the large $M$ limit of (\ref{infdensm}) yields
\be
\rho(x,y)=\frac{1}{\pi} \frac{1}{z} \partial_{\bar{z}}G(z,\bar{z})
\label{infdens}
\ee
As a cross-check of our calculations we verified that the spectral function
obtained in this way is real.

Substituting $b^2=0$ in (\ref{finalgap}), we get the equation for the shape 
of the support of the eigenvalues for arbitrary $\tau$. Note
that the shape of the boundary depends on the variable $\log |z|$, reflecting
in the spectrum the multiplicative (geometric) feature of the 
stochastic process induced by the multiplication of random
matrices.
Explicit solution for the   boundary reads
\be
\frac{\tau}{2}\left( \frac{r^2-1}{\ln r}\right) = r^2 +1 -2r \cos \phi
\label{conformalcurve}
\ee
It is rewarding to compare solution (\ref{conformalcurve}) to the
solution
of the conchoid-like boundary in the case of two complex matrices 
(\ref{conchoid}). In the limit of infinitely many products, 
the shape of the boundary acquires a new symmetry --
(\ref{conformalcurve}) is invariant under inversion operation $ r \rightarrow
1/r$.
It is visible also from the general form (\ref{finalgap}), since the 
radial dependence  for $b^2=0$  is a function of $|\log r |$ only. 
This symmetry  is responsible for the appearance of a structural change of
the spectrum -- provided  $\tau$ is sufficiently large, {\em two}
boundaries, related by inversion in the radial variable, appear.  
%For too small $\tau$, one of the twinned boundaries has no solution 
%for $|\cos \phi | \le 1$. 
We describe this structural change of the
complex  spectrum as a "topological phase transition". 
In Section~8, we  study numerically the spectral density 
predicted by our formulas, in particular,  
the ``diffusion'' of the shape of the boundary 
caused by evolution with  the ``time'' parameter $\tau$. 
We  also 
confirm  numerically  the  critical behavior  for certain 
$\tau_{\rm crit}$.

\section{Product of Two Hermitian Matrices}

In this section we will consider the spectral distribution
of the product 
\be
Y_2=
\left(1+\sqrt{\frac{\tau}{2}}H_1\right) \left(
1+\sqrt{\frac{\tau}{2}}H_2\right) 
\label{case2h}
\ee
where $H_1, H_2$ are {\em not} arbitrary complex matrices, 
but they are identical independent hermitian Gaussian matrices (iiGUE). 
Contrary to naive expectations,
the above problem is more subtle then the analogous problem of the two 
complex Gaussian matrices. 
First, let us note, that the product of two hermitian matrices is 
{ \em not}, in general, a hermitian matrix, so the spectrum of the 
strings of hermitian matrices develops complex eigenvalues. 

Second, the case of only two hermitian matrices of the considered type 
turns out to be in  some sense singular -- 
for certain values of the evolution 
parameter $\tau$, the {\it a priori} complex eigenvalues get localized
on the real axis. This phenomenon reminds a bit the non-hermitian localization 
in the Hatano-Nelson model~\cite{HATANONELSON}. 
Third, the problem is algebraically more involved since there will be
more nonzero propagators than in the complex case.

Since even a product of two hermitian matrices can develop {\it a priori}
a complex spectrum,
we have to use the same type of auxiliary Green's function as in the 
complex Gaussian case. 
The construction parallels exactly the scheme for two complex 
matrices considered previously, with the matrices $X_i$ and
$X_i^\dagger$ replaced both by $H_i$. The first new feature appears when 
we construct the ``Schwinger-Dyson'' equations for the 4 by 4 
self-energy matrix $\tilde{\Sigma}$. Since the matrices 
are hermitian ($H_i=H_i^\dagger$), additional nonzero contractions
appear in the matrix $\tilde{\Sigma}$
\be
\tilde{\Sigma} = \alpha \left(  
\begin{array}{cccc} 0&g_{12} &g_{1\bar{1}}& 0  \\ 
g_{21}&0 &0 &g_{2\bar{2}} \\ 
g_{\bar{1}1}&0&0 &g_{\bar{1}\bar{2}} \\ 
0&g_{\bar{2}2} &g_{\bar{2}\bar{1}}&0 \\
\end{array}\right)
\equiv \alpha \left(  
\begin{array}{cccc} 0&h &g& 0  \\ 
h&0 &0 &g \\ 
\tilde{g}&0&0 &\tilde{h} \\ 
0&\tilde{g} &\tilde{h}&0 \\
\end{array}\right)
\label{hermiteancase2}
\ee
where again the factor $\alpha=\tau/2$ comes from the propagator, $h$
and $\tilde{h}$ are the new entries and
we have exploited the symmetries of the problem (e.g. $
g_{1\bar{1}}=g_{2\bar{2}}$ etc).

The solution of the 4 by 4 matrix equation for the 4 by 4 Green's function
\be
\GG(w)=({\cal{W}} - \tilde{\Sigma})^{-1}
\ee
is more involved, since we have to  solve it 
not only for the ``gaps''  $g$ and $\tilde{g}$ 
but also for additional gaps  $h$ and $\tilde{h}$. 
Only then  we can calculate $g_{11}$, 
to recover the wanted eigenvalue distribution in the same way as in
the complex case (\ref{e.rhotwo}).

%$G(z,\bar{z})=\frac{g_{11}(w)}{w}$.

\begin{figure}[t]
{\epsfysize=30mm 
\hfill\epsfbox{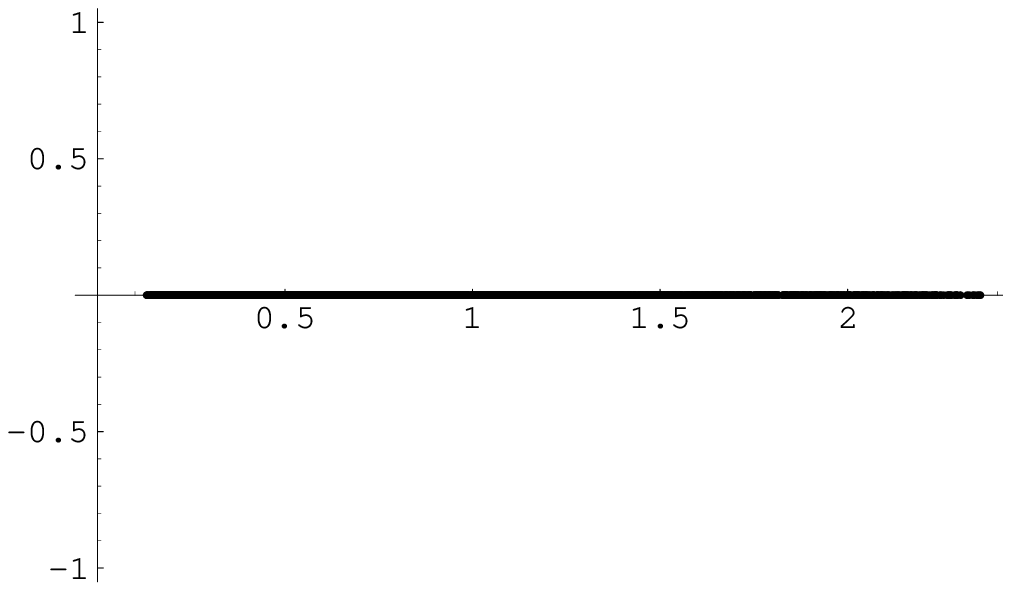}\hfill%
\epsfysize=30mm \hfill\epsfbox{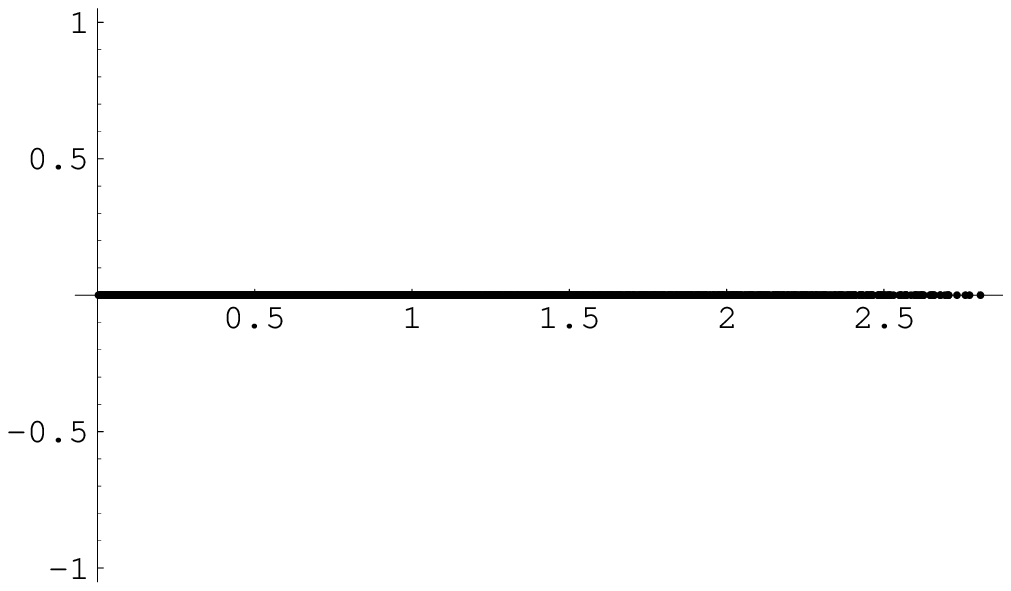}\hfill\mbox{ }}
\caption{Complex eigenvalues of 10  products of two $200\times 200$ hermitian
matrices before and at the transition ($\sqrt{\tau/2}=0.4$ and $0.5$).
The $x$ and $y$ axis correspond to $\Re \lambda$ and $\Im \lambda$, 
respectively.}
\label{fig.transitiona}
\end{figure}

\begin{figure}[t]
{
\mbox{ }\epsfysize=30mm\hfill\epsfbox{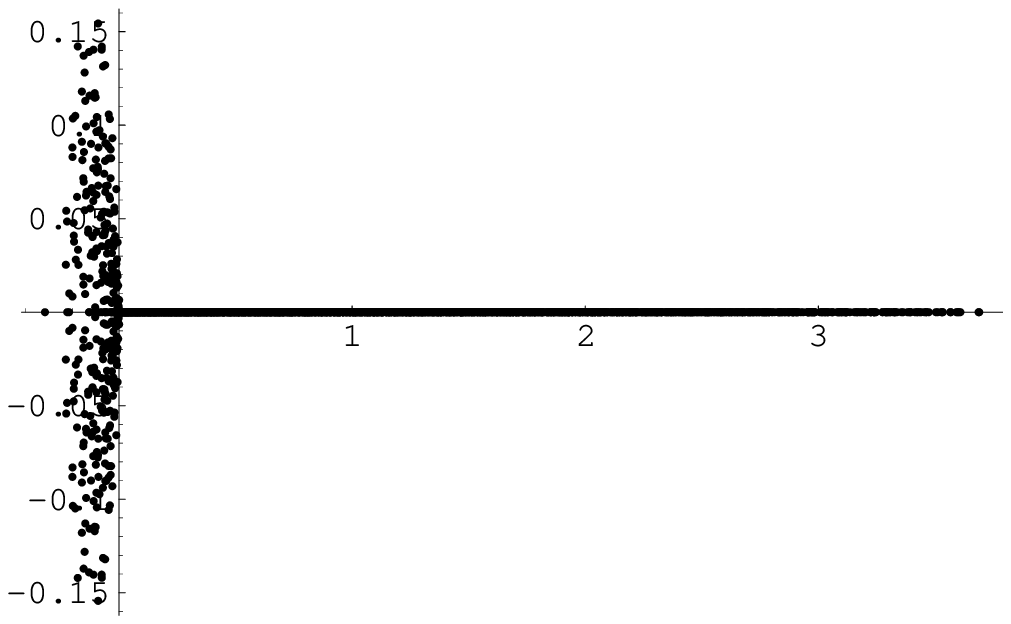}\hfill
\epsfysize=30mm\hfill\epsfbox{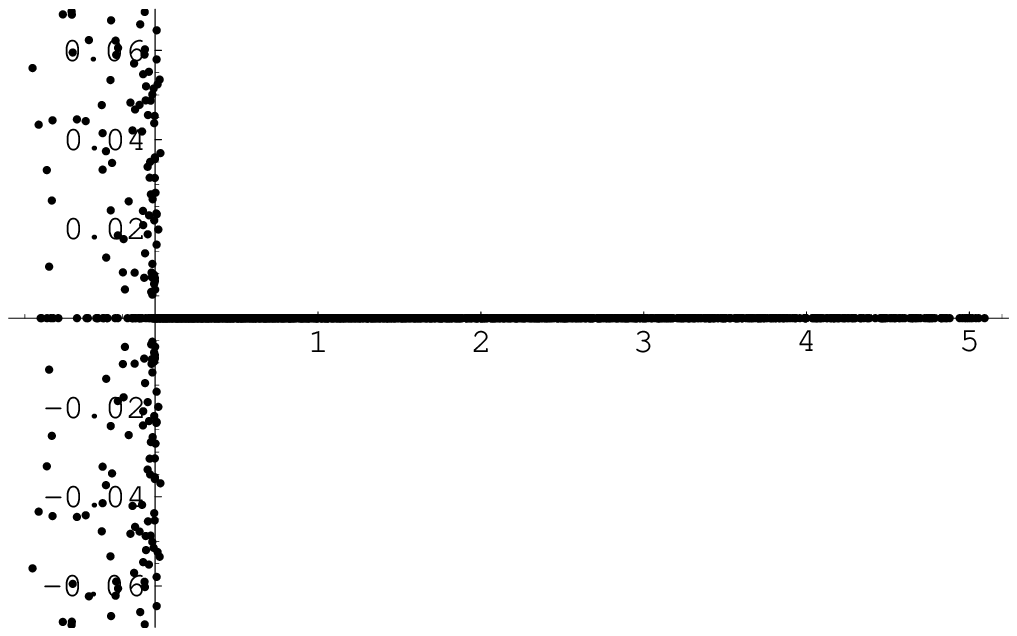}\hfill}

\caption{Eigenvalues of 10 products of two $200\times
200$ hermitian matrices after the transition ($\sqrt{\tau/2}=0.7$ and
$1.0$). The axis are defined as in the previous figure. } 
\label{fig.transitionb}
\end{figure}

Introducing $d^2=(1+\alpha h )(1+\alpha \tilde{h})$ and $b^2=g\tilde{g}$,
we obtain two (modulo their tilded  partners) gap equations
\be
g &=& -\frac{\alpha g}{det} (\alpha^2 b^2  -|w|^2 - d^2)
\nonumber \\
h &=& \frac{1}{det} (\alpha^2 b^2 (1+\alpha \tilde{h}) + 
(1+\alpha h)(\bar{w}^2-(1+\alpha \tilde{h})^2)
\label{e.gh}
\ee
coupled to the needed $g_{11}$
\be
g_{11}= \frac{1}{det} 
(-\alpha^2 b^2 \bar{w} +w(\bar{w}^2-(1+\alpha \tilde{h})^2)
\ee
where $det$ is the determinant of the matrix ${\cal W} -\tilde{\Sigma}$
 and reads explicitly
\be 
\label{e.hdet}
det=\alpha^4 b^4 +(w^2-(1+\alpha h)^2)(\bar{w}^2 -(1+\alpha \tilde{h})^2)
-2\alpha^2b^2 (d^2 +|w|^2)
\ee
Numerical simulations show that the eigenvalue distribution for the
product of two matrices is concentrated on an interval on the {\em
real} axis for $\tau<1/2$ and only then when the edge of the interval
reaches the origin, do the eigenvalues start to spread into the
complex plane (see figures \ref{fig.transitiona} and
\ref{fig.transitionb}). 
Let us show how to find the critical value $\tau=1/2$ from the above
formulas based on the above numerical observation. 

We thus have to find the point when the boundary of the two dimensional
island hits the origin $w=0$. Then the equations (\ref{e.gh}) and
(\ref{e.hdet}) simplify to
\eqn
h &=& \f{-1}{1+\al h} \\
1 &=& \al h \tilde{h}
\eqnx
where we set $b^2=g \tilde{g}=0$. Then it is easy to verify that this
set of equations is satisfied for $\al=1/4$ which is equivalent to
$\tau=1/2$.

\section{Products of Arbitrary Number of Hermitian Matrices}

Finally, we  consider an arbitrary number of matrices 
belonging to iiGUE.
\be
Y_M=
\left(1+\sqrt{\frac{\tau}{M}}H_1\right) \left(
1+\sqrt{\frac{\tau}{M}}H_2\right)  \dots \left(
1+\sqrt{\frac{\tau}{M}}H_M\right) 
\label{caseMh}
\ee
As before, the construction of the auxiliary Green's function
parallels the complex case, up to the point when we have to construct
explicitly the matrix of self-energies $\tilde{\Sigma}$, corresponding
to the set of $4M^2$  Schwinger-Dyson equations. 
Generalization of the results from the previous sections leads to the 
explicit form of the auxiliary Green's function. 
Introducing 
\be
{\cal G}(w)=   \left( 
\begin{array}{cc} {{\cal W}}&{\cal S} \\ 
{\tilde{{\cal S}}}& {\cal W}^{\dagger}
\end{array}\right)^{-1}  
\ee
with blocks 
\be 
{\cal W}= \left( 
\begin{array}{ccccc} 
 w  & - 1-\alpha h  & 0 & \dots & 0 \\
   0     &   w & - 1-\alpha h& \dots & 0\\
  \dots & \dots & \ldots & \ldots & \ldots \\
  0& 0& \ldots &  w & -1-\alpha h \\
   -1-\alpha h & 0 & \ldots & 0 &  w
\end{array} 
 \right)_{M \times M}   
\ee
and 
\be 
{\cal S}= - \alpha \left( 
\begin{array}{ccccc} 
g & 0& 0 & \dots & 0 \\
   0     &  g & 0 & \dots & 0\\
  \dots & \dots & \ldots & \ldots & \ldots \\
  0& 0& \ldots & g & 0 \\
  0 & 0 & \ldots & 0 & g
\end{array} 
 \right)_{M \times M}   
\ee
the consistent set of equations comes 
as 
\be
g_{11}&=&{\cal G}_{1,1} \nonumber \\
h&=&{\cal G}_{1,M} \nonumber \\
\bar{h}&=&{\cal G}_{M+1,M+2} \nonumber \\
g&=&{\cal G}_{1,M+1} \nonumber \\ 
\bar{g}&=&{\cal G}_{M+1,1}
\ee
i.e. requires an explicit inversion of the $2M$ by $2M$ matrix. 
The algebra is more complicated since, like in the case of two hermitian 
matrices, we have now two types of gap equations (for $g$ and $h$  and their
tilded partners).
Using the solutions, we calculate $g_{11}$, 
then, 
%we recover the original function 
% $G(z,\bar{z})=\frac{g_{11}(w)}{w^{M-1}}$
we get an explicit equation for the spectral density (\ref{densref}).
%\be
%\rho(x,y)=\frac{1}{\pi}\partial_{\bar{z}}G(z,\bar{z})
%\ee
Finally, we perform the limiting procedure.
Technically, one has to repeat the transfer matrix technique 
introduced in section~5 for  the complex case.  
The algebra is lengthy, but due to the cyclic nature of the blocks 
of matrix  ${\cal R}$ the results form a rather surprisingly simple set 
of equations. Returning to the original variable $z=r\exp (i\phi)$ 
%and
%introducing $h\equiv A+iB$ 
 we get
\be
 G(z, \bar {z})= \frac{1}{2\tau} \log r +\frac{3}{4}  -\frac{iU}{\tau}
\sin{\psi} / \sinh U
\label{finalhermit}
\ee
and the gap equations read  
\be
h= \frac{1}{2\tau} \log r -\frac{1}{4} -\frac{iU}{\tau}
\sin{\psi} / \sinh U = G-1
\label{finalhermith}
\ee
and 
\be 
\cos \psi= \cosh U -\frac{\tau}{2U} \sinh U
\label{finalgaph}
\ee
where  $\psi=\phi+U \frac{\sin \psi}{ \sinh U}$
and  
$U=\sqrt{(\frac{\log r}{2} + \tau/4)^2 -\tau^2 g\tilde{g}}
$.
The above set of equations completes 
the solution of the continuous product of matrices  with Gaussian hermitian
disorder. Spectral density follows from (\ref{infdens}). 
Note, that the dependence on  $\log|z|$  reflects
 the multiplicative
character of the Brownian motion. Again, the boundary  
has also an inversion-type symmetry, but now it is 
of the form $|z| \rightarrow \exp(-\tau)/|z|$. 
Due to the non-zero gap $h$, the boundary in the hermitian case is
given
by the pair of coupled transcendental equations and can be presented
numerically only.   The appearance 
of the aforementioned  symmetry is responsible for the "topological phase transition"
in the spectrum of the product of infinitely many matrices,
similar to the one observed for the product of infinitely many complex
matrices.

In Section~8, we  study numerically the spectral density 
predicted by our formulas, in particular,  
the ``diffusion'' of the shape of the boundary 
caused by evolution with  the ``time'' parameter $\tau$. 
We confirm the appearance of a ``phase transition'' 
for certain $\tau_{\rm crit}$ in the hermitian case as well.    

\section{Numerical Tests}

\begin{figure}[htbp]
%%%%%%%%%%%%% complexevolution graph graph %%%%%%%%%%%%%%%%%%%%%%
\centerline{\epsfysize=55mm \epsfbox{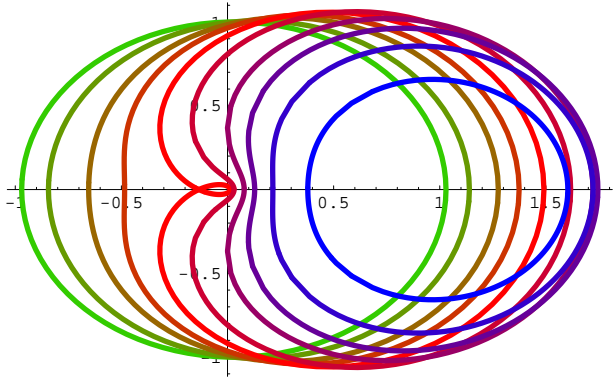}}
%%%%%%%%%%%%% 1st graph %%%%%%%%%%%%%%%%%%%%%%
\caption{Evolution of the rescaled (see text) 
contour of non-holomorphic domain 
in the eigenvalue spectrum of the %product of infinitely many Ginibre-Girko
%ensembles
Ginibre-Girko multiplicative diffusion
as a function of several times $\tau$.
}
\label{fig.comev}
\end{figure}

\begin{figure}[htbp]
%%%%%%%%%%%%% tau4 %%%%%%%%%%%%%%%%%%%%%%
\centerline{\epsfysize=55mm \epsfbox{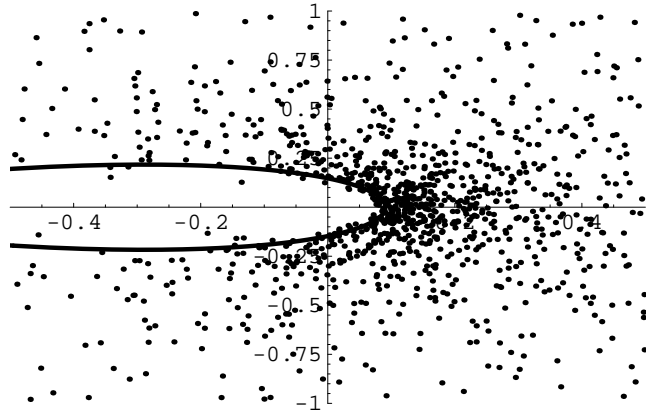}}
%%%%%%%%%%%%% 1st graph %%%%%%%%%%%%%%%%%%%%%%
\caption{Comparison of the analytical contour for 
the eigenvalue spectrum of %product of infinitely many 
%ensembles
Ginibre-Girko multiplicative diffusion 
at evolution time $\tau=4.$, versus the high-statistics numerical simulation
of the spectrum.}
\label{tau4hole}
\end{figure}

\begin{figure}[htbp]
%%%%%%%%%%%%% dziura %%%%%%%%%%%%%%%%%%%%%%
\centerline{\epsfysize=55mm \epsfbox{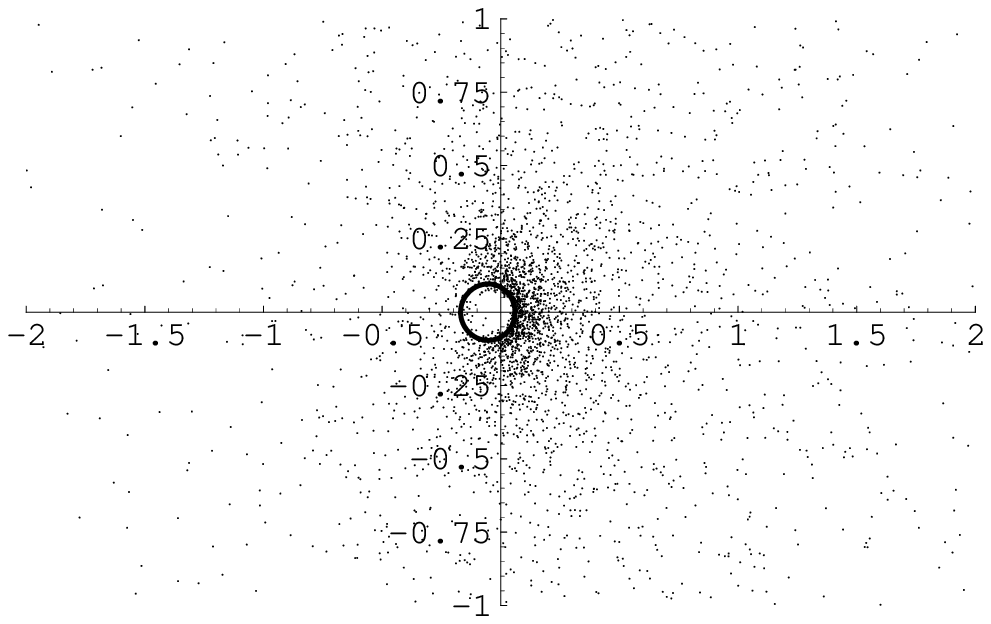}}
%%%%%%%%%%%%% 1st graph %%%%%%%%%%%%%%%%%%%%%%
\caption{Comparison of the analytical contour 
for the eigenvalue spectrum of %product of infinitely many 
%ensembles
Ginibre-Girko multiplicative diffusion 
at evolution time $\tau=5$, versus the numerical simulation
of the spectrum. The presence of few eigenvalues inside the inner
contour
is caused by numerics, i.e. finite $N$ and finite $M$ effects.}
\label{tau5hole}
\end{figure}

In this section we present the results of numerical simulations,
confirming
our analytical predictions. 
We start from the complex case. Figure~\ref{fig.comev} shows
the evolution of the boundary for several sample evolution times
$\tau=0.5,1,1.5,2,3,4,5,6,8,12$. To present the whole set of the
boundaries
on the same figure, each boundary is rescaled by a corresponding factor
$\exp(-\tau/2)$. We would like to note here that the effect of 
rescaling is equivalent to the non-rescaled process
with drift $\mu={\bf 1}$. One can see how the original 
ellipse-like shape (innermost figure) evolves through  a twisted-like
shape to the set of double-ring structures. 
The inner ring, always containing the origin, is so small on the scale
of the Fig.~\ref{fig.comev} that  is not visible.
At $\tau=4$, corresponding to the curve with an inner loop, we observe
a topological phase transition. The support of the spectrum is no longer
simply connected, for $\tau \ge 4$ it is annulus-like, 
i.e. eigenvalues are expelled from the central region. 
For even larger times, the outside rim of the annulus approaches the
circle,
and the inner one shrinks to the point $z=0$ in the $\tau=\infty$
limit.
Indeed for large $\tau$ the radius of the inner ring behaves like
\eq
r(\tau) \sim e^{-\f{\tau}{2}}
\eqx
The outer boundary then forms approximately a circle with the radius
$e^{\tau/2}$.
To visualize the repulsion of the eigenvalues from the region around $z=0$ 
we performed  higher statistics simulations for $\tau=4$ and $\tau=5$, 
corresponding to Figs~\ref{tau4hole},\ref{tau5hole},
respectively. Note that the eigenvalue distributions are quite high
for small $z$ and the size of the inner ring is indeed very
small. Therefore it is quite difficult to observe numerically the
exact exclusion of eigenvalues from the marked circle.

\begin{figure}[htbp]
%%%%%%%%%%%%% c1.0 %%%%%%%%%%%%%%%%%%%%%%
\centerline{\epsfysize=55mm \epsfbox{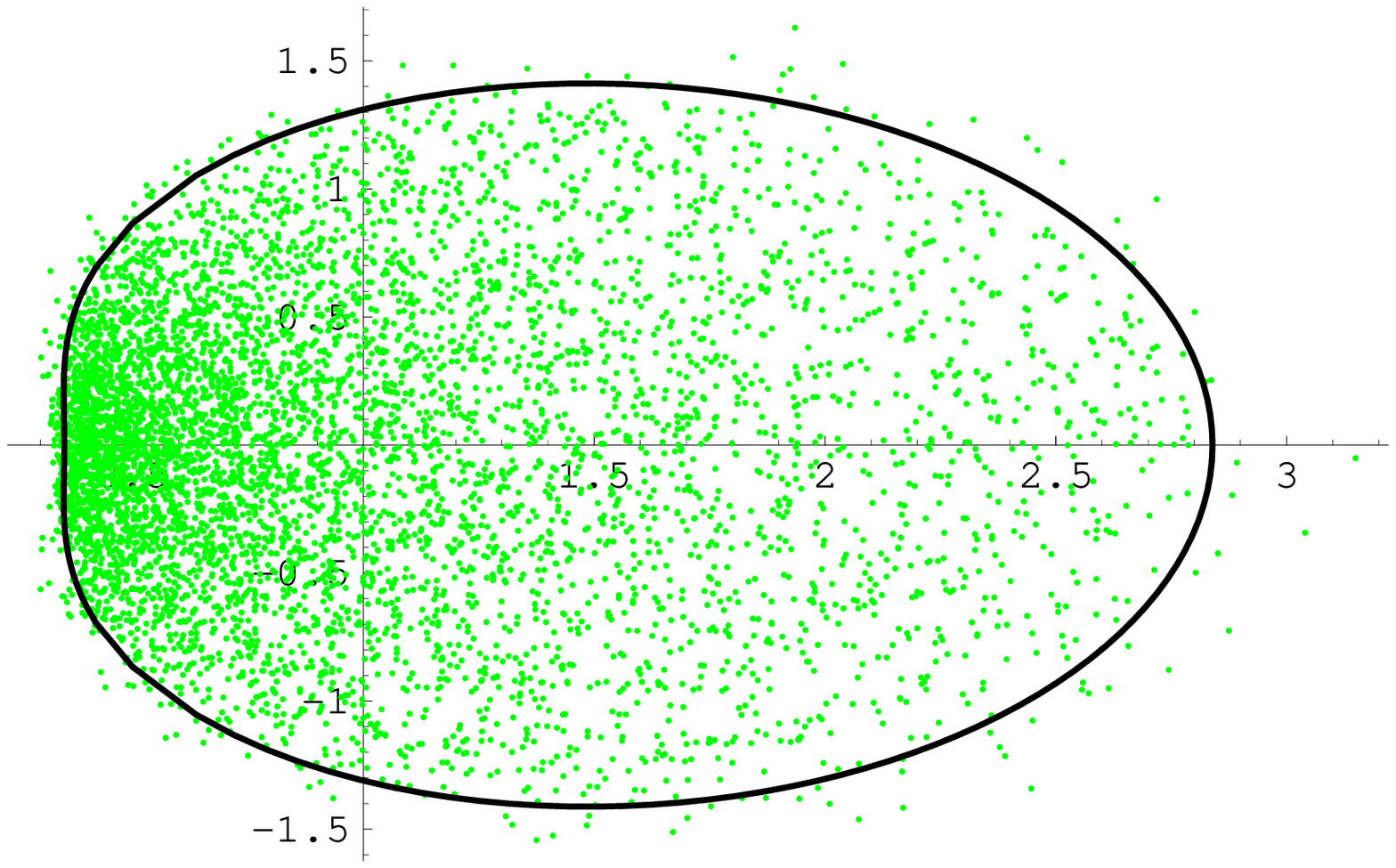}}
%%%%%%%%%%%%% 1st graph %%%%%%%%%%%%%%%%%%%%%%
\caption{Comparison of the analytical contour 
for the eigenvalue spectrum of %product of infinitely many  Ginibre-Girko
%ensembles
Ginibre-Girko multiplicative diffusion 
at evolution time $\tau=1.$, versus the numerical simulation
of the spectrum. The generated ensemble consisted of 60 matrices,
each for $N=M=100$. Note that the vertical axis is located at $x=1$
and not at $x=0$. The origin lies outside the figure.}
\label{fig.com1}
\end{figure}

Figure~\ref{fig.com1} shows the comparison of numerically
generated spectra versus the analytical prediction of the shape 
of the support of  eigenvalues of the "evolution operator".
Again, the same rescaling by $\exp(-\tau/2)$ was applied.

\begin{figure}[htbp]
%%%%%%%%%%%%% hermitevolutiongraph%%%%%%%%%%%%%%%%%%%%%%
\centerline{\epsfysize=55mm \epsfbox{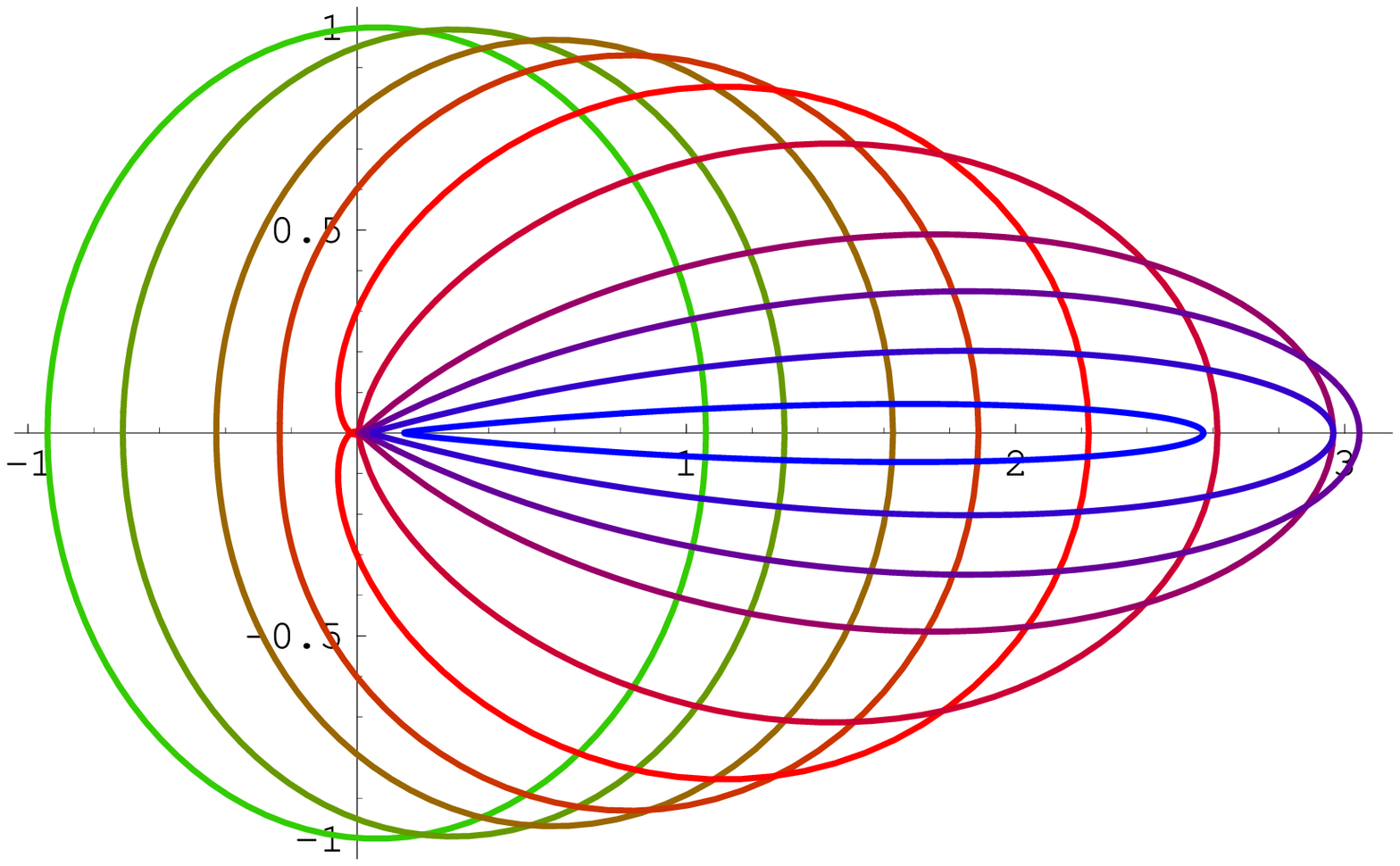}}
%%%%%%%%%%%%% 1st graph %%%%%%%%%%%%%%%%%%%%%%
\caption{Evolution of the contour of non-holomorphic domain 
in the eigenvalue spectrum of the %infinite product of GUE ensembles
GUE multiplicative diffusion
as a function of evolution times $\tau=0.5,1,1.5, 2,3,4,5,6,8,12$,
 from inside to outside, 
respectively.}
\label{fig.hermev}
\end{figure}

Figure~\ref{fig.hermev} shows
the evolution of the boundary for several sample evolution times
$\tau=0.5,1,1.5,2,3,4,5,6,8,12$ in the case of GUE. 
Again, to present the whole set of
boundaries
on the same figure, each boundary was rescaled by a corresponding factor
$\exp(-\tau/2)$. 
One can  observe how the almost real (for small times $\tau$)
long-cigar shaped spectrum evolves into a broader shape, 
developing again singular behavior at $\tau=4$ at the leftmost-edge
of the spectrum. Again, the spectrum develops a topological
transition,
although it is not as clearly visible in  the figure as in the complex
case.
The reason is that inversion symmetry has an additional exponential
suppression factor, i.e. $r \rightarrow e^{-\tau}/r$.
For very  large times, the outside rim of the spectrum approaches the
circle,
and the inner one shrinks to a point in the $\tau \to \infty$ limit.

\begin{figure}[htbp]
%%%%%%%%%%%%% h1.0 %%%%%%%%%%%%%%%%%%%%%%
\centerline{\epsfysize=55mm \epsfbox{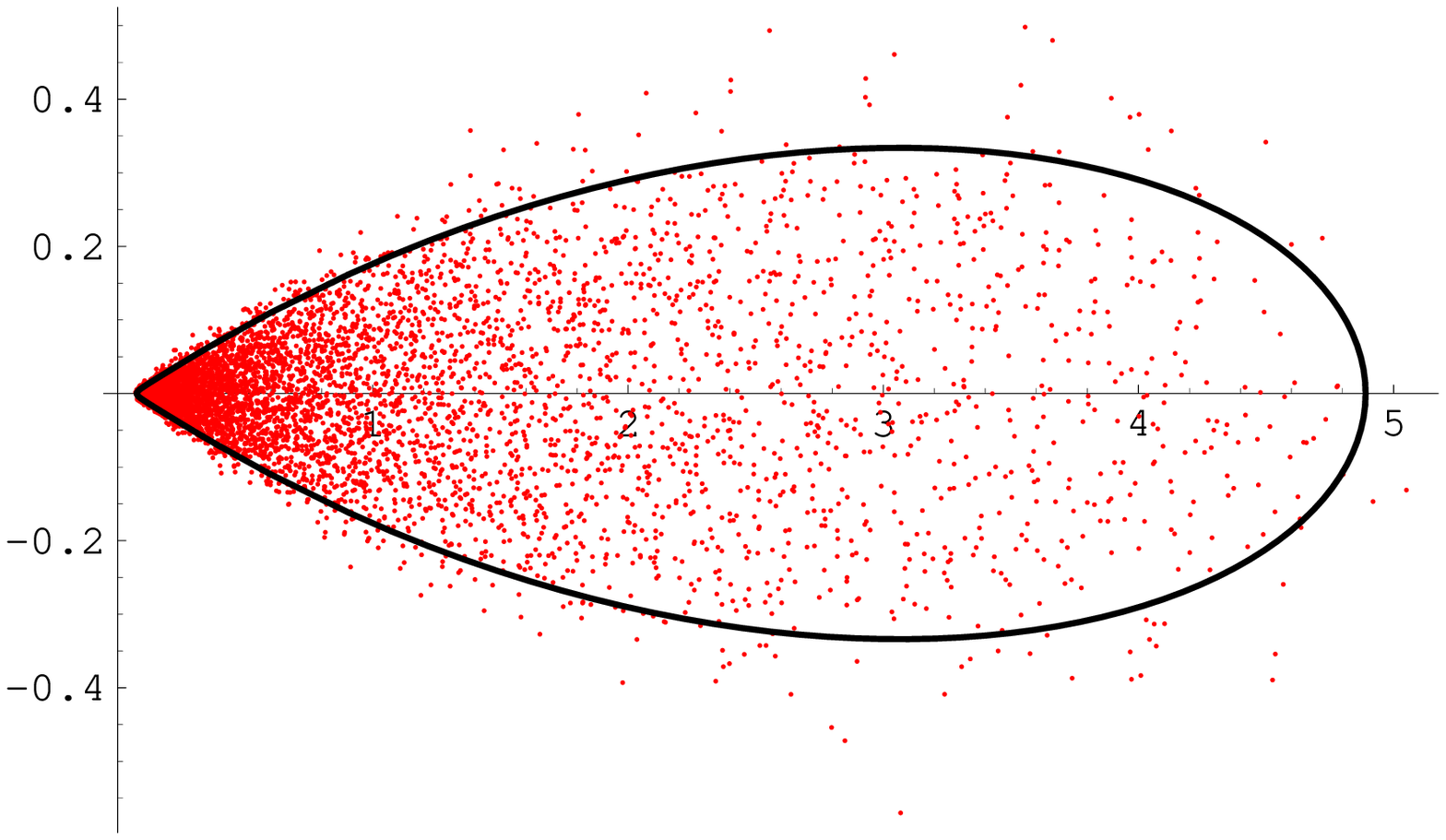}}
%%%%%%%%%%%%% 1st graph %%%%%%%%%%%%%%%%%%%%%%
\caption{Comparison of the analytical contour 
for the 
the eigenvalue spectrum of %product of infinitely many  GUE's
GUE multiplicative diffusion
at evolution time $\tau=1.$, versus the numerical 
simulation of the spectrum. The ensemble consists of 60 matrices, 
each obtained as a string of $M=100$ of $N$ by $N$ matrices, where $N=100$. }
\label{fig.her1}
\end{figure}

As in the complex case, here we also show a sample 
simulation~(Fig.~\ref{fig.her1}) versus 
the analytical prediction for the shape of the island.

Finally in figures \ref{fig.transitiona} and \ref{fig.transitionb} above, we
demonstrated the "localization-delocalization" 
phase transition in the case of the product of two hermitian matrix
models of the considered type. Below $\tau_{\rm crit}$, 
a priori complex spectra condense  on the positive real axes. 
At $\tau=1/2$, the eigenvalues start to diffuse onto the $x \le 0$ 
half-plane. For very large times  $\tau$, 
eigenvalues start to appear on both sides of the $x=0$ axis.

\section{Stability and Lyapunov Exponents}

In this section, we briefly discuss the relation between the spectrum of the
``evolution operator'' $Y_M$ and Lyapunov spectra.
In the studies of products of random matrices one of the central issues is the 
use of limit theorems that would provide an information about the 
asymptotic limit of the rates of growth and of the spectrum of the
product for a sequence of matrices. In particular the limit
\be
\lim_{M\rightarrow\infty}\frac{1}{M}\left\langle \ln |Y_M|
\right\rangle \equiv \lambda_1
\ee
defines the  maximum Lyapunov characteristic exponent. The existence
of this limit is assured for an infinite product of matrices by the
Furstenberg~\cite{VULPIANI} theorem   
\be
Y_M=\prod^{M}_{i=1} X(i)
\ee
for independent random matrices $X(i)$ characterized by the probability
distribution function $d\mu[X]={\cal P}[X]d[X]$. It can be shown under very
general assumptions that $\lambda_1$ is a non-random, self-averaging quantity,
so that the brackets $\cor{\ldots}$ 
in the above expression can be dropped in almost any
realization. The best known example is the use of $\lambda_1$ 
in the description
of the DC conductivity of a one dimensional disordered system coupled to two
electron reservoirs. The conductivity $\kappa$ of the system, 
given by the Landauer
formula
%\be
%\kappa=\frac{q_e^2}{2\pi\hbar}\frac{|t|^2}{|r|^2}
%\ee
%where $q_e$ is the electron charge and $t,r$ stand for the transmission and
%reflection coefficients, respectively. The transmission coefficient is related
%to the product $Y_M$ of the transfer matrix $Y_M= X(M-1)X(M-2)...X(0)$:
%\be 
%|t|=\frac{2|\sin
%k|}{|(Y_M)_{21}-(Y_M)_{12}+(Y_M)_{22}e^{ik}-(Y_M)_{21}e^{-ik}|} 
%\ee
yields for large $M$ 
\be
\kappa \approx e^{-2\lambda_1 M}
\ee
where $\lambda_1$ is the maximal Lyapunov exponent of the product $Y_M$.
By the Borland conjecture, the inverse of $\lambda_1$ gives the
localization length measuring the decay of an eigenvector of the
system's Hamiltonian over the chain composed of $M$ units. 

It turns out that the formalism of  generalized Green functions
(Section 2 and 3) is particularly suitable for deriving an analytical formula
\cite{EWA1,EWA2} for $\lambda_1^{-1}$ for the one-dimensional
disordered wires within the tight-binding 
Hamiltonian with a site diagonal disorder. Extensions to generalized
Lyapunov exponents $L(q)$ \cite{VULPIANI} related to the exponential
growth rates of the moments of the matrix product 
\be
L(q)=\lim_{M\rightarrow\infty}\frac{1}{M} \ln \left<  |Y_M|^q \right>
\ee
have been also 
reported for systems described by Markovian  evolutionary operators
\cite{OLIVEIRA} used in the study of DC conductance statistics of
the random dimer model introduced for explaining the exceptionally
high electronic transport properties of some conjugated polymer
chains. 

The eigenvectors of the product $Y_M^{\dagger}Y_M$ 
form an orthonormal set of vectors  (Lyapunov
basis) corresponding to characteristic Lyapunov exponents whose  exponentials,
$\exp(\lambda_i)$ are eigenvalues of the matrix
$|Y_M^{\dagger}Y_M|^{1/2M}$ in the limit of large $M$ (Oseledec's theorem).

The Lyapunov exponents are closely related to the linear stability analysis 
of the dynamic system $\dot{x}=F(x)$ for which small perturbations $y$ around
a given solution $x(t)$ evolve in time according to $y(t_2)=K(t_2,t_1)y(t_1)$.
They are defined as
\be
\lambda_i=\lim_{t_2\rightarrow\infty}\frac{1}{t_2-t_1} \log
||K(t_2,t_1)f_i(t_1)||
\ee
and for ergodic systems their spectrum  describes global properties  of the
system's attractor (does not depend on the initial point).
In contrast to the Lyapunov spectra, the Lyapunov basis  for the
ergodic sequence of matrices describes local properties of the PRM system.

Here we would like to underline, that Lyapunov exponents are in
general different from the spectrum of eigenvalues obtained in this 
paper. Since the first Lyapunov exponent is defined as a property 
of the {\em norm} of the operator $Y$, it is related 
to the spectral properties of the $Y^{\dagger}Y$ operator. 
This is why the Lyapunov exponents are always real.
The spectra studied in this paper correspond to so-called
stability exponents, which are directly related to the eigenvalues of
the $Y$ operator itself.  
As we have demonstrated above, the eigenvalues of $Y_M$  are in
general complex numbers and so are the stability exponents: 
\be
\alpha_i=\lim_{M\rightarrow\infty}\frac{1}{M}\left< \ln \lambda_i\right>
\ee
Their real parts coincide with the Lyapunov exponents
$\lambda_i=Re(\alpha_i)$ only if the matrix $Y_M$ can be written in a
diagonal form.
 
The relation between the stability and Lyapunov exponents in a broad
class of dynamical systems has been investigated by Goldhirsch {\it
et al.} \cite{GOLDHIRSCH}. Motivated by mostly numerical results, 
the authors have conjectured that Lyapunov exponents equal stability
ones for the ergodic dynamical systems whose spectrum is
non degenerate and the attractor is bounded. The geometrical argument
behind that statement is that generically, stretching and shrinking
of the relative separation of ``trajectories'', starting at a given
initial point happen along fixed directions in the phase space. 
Taken from that perspective further elucidation would be required to
clarify the relation between the Lyapunov stability and spectral properties of
the generalized stability matrix  investigated in this work.

\section{Summary}

We have introduced a natural generalization of the  concept of
geometric random walk (`geometric diffusion') in the space of 
large, non-commuting matrices.
In particular, we proposed a simple diagrammatic method, allowing to 
calculate spectral properties of the resulting evolution operators as
functions of the "evolution" time $\tau$. The construction was
explicitly presented in the 
case of complex Gaussian ensembles and  GUE ensembles. 

Our method was based on an exact relation between the eigenvalues
of a product matrix and the eigenvalues of a certain block matrix.
The essential simplification is the fact that the random matrices
appear {\em linearly} in the block matrix.
Although we used the above identification only in the large $N$ limit, 
it could a-priori be used to study the properties of the spectrum
of the product matrix also in the microscopic limit, i.e. on the scale
of eigenvalue spacing. But then different, non-diagrammatic, methods 
would have to be employed.

Despite the 
simplicity of the random matrix models chosen as the building blocks 
of the matrix-valued diffusion, the spectrum develops a 
surprising feature, namely a region {\em without} eigenvalues appears
within the spectrum, thus changing its topological properties.
We can thus describe this behavior as a topological phase
transition. This points at the appearance of particular 
{\it repulsion} mechanism for sufficiently large evolution times.
Our results were based on the simulation based on Gaussian ensembles.
We are however tempted to conjecture, that unfolding the spectrum in the
vicinity of the point where the topological change occurs may  unravel a
possible novel {\em universal} spectral behavior.
We would like to mention, that the critical behavior described here 
as a "topological phase transition" appeared only in the limit
$M \rightarrow
\infty$, as a consequence of the inversion symmetry, acquired in this limit, 
of the equation
describing the boundary of the two-dimensional support of
eigenvalues.   
We would like to mention, that similar, qualitative effect of
expulsion of complex eigenvalues from the certain region of complex
plane
was observed by Feinberg and Zee (first reference
in~\cite{OTHERNONHERM}). They considered however, a single random
complex ensemble with non-Gaussian measure and they imposed the rotational
symmetry of the spectrum. In our case, we considered the infinite
product representing the Ginibre-Girko (therefore Gaussian) 
multiplicative diffusions, and we obtained the structural change of
the spectrum for the general case, with non-trivial dependence on the
polar variables. It is an intriguing problem for  further
studies, to what  extent these results reflect the general  
"single ring hypothesis".

    We pointed out, in agreement
with~\cite{GOLDHIRSCH}, that stability exponents considered in our
paper carry much richer information about  the properties of the system 
than the Lyapunov exponents.
     
As a by-product of our analysis, we found a previously unknown (as far
as we know)       
critical behavior in the case of the particular product of two hermitian
ensembles. The {\em a-priori} complex spectra of the product are
nevertheless real for evolution times below a certain critical value
$\tau_{crit}$, and migrate into the complex plane only for
$\tau>\tau_{crit}$.  
A similar in spirit localization-delocalization 
phase transition was recently observed 
for a class of non-hermitian Anderson Hamiltonians~\cite{HATANONELSON}.   

One of the motivations for this work was to propose  a general 
formalism, 
which can provide a straightforward  method of analyzing spectral properties
of multivariate diffusion-like processes, with the idea that 
our method could be used in different branches of theoretical physics
and interdisciplinary applications. 
Some particular applications will be presented elsewhere. 
Finally, we would like to mention, that the presented formalism allows
also to establish a direct link to the diffusion processes
based on Free Random Variables techniques~\cite{VOICULESCU,SPEICHER,BIANE}, in
particular it can demonstrate
the emergence of the complex Burgers equations governing the
evolution of the spectral generating functions~\cite{WIECZETAL}.

\subsubsection*{Acknowledgments}
This work was partially 
supported by the Polish State Committee for Scientific Research
(KBN) grants 2P03B 09622 (2002-2004), 2P03B08225 (2003-2006) and the EU
Center of Excellence 
in Information Society Technologies "COPIRA". 
JJ and MAN would like to thank the Niels Bohr Institute,
where a part of this work has been completed, for the hospitality.
Their stay at NBI was
supported by ``MaPhySto'',
the Center of Mathematical Physics
and Stochastics, financed by the
National Danish Research Foundation.
 \\
 The authors would like
to thank Angelo  Vulpiani for interesting discussions.
We are also very grateful to Roland Speicher and Piotr \'Sniady for comments
after submitting this paper to {\tt math-ph} and 
for valuable remarks 
in relation to Free Random Variable calculus.

%\vglue 2.5cm

%%
%%Close Feynman

\end{document}